\pdfoutput=1 
\documentclass[12pt,letterpaper,oneside]{article}
\usepackage{caption}
\usepackage[]{natbib}
\usepackage{graphicx}
\usepackage{subcaption}
\usepackage{ulem}
\usepackage{amsmath}
\usepackage[margin=1in]{geometry}
\usepackage{xcolor}
\usepackage{setspace}
\usepackage{float}

\usepackage{authblk}
\title{Extrapolated full waveform inversion with deep learning}
\author{Hongyu Sun and Laurent Demanet}
\affil{Massachusetts Institute of Technology, 77 Massachusetts Ave, Cambridge, MA 02139, hongyus@mit.edu, laurent@math.mit.edu}
\date{April, 2019}

\begin{document}

\maketitle

\begin{abstract}
The lack of low frequency information and a good initial model can seriously affect the success of full waveform inversion (FWI), due to the inherent cycle skipping problem. Computational low frequency extrapolation is in principle the most direct way to address this issue. By considering bandwidth extension as a regression problem in machine learning, we propose an architecture of convolutional neural network (CNN) to automatically extrapolate the missing low frequencies without seismic preprocessing and post-processing steps. The bandlimited recordings are the inputs of the CNN and, in our numerical experiments, a neural network trained from enough samples can predict a reasonable approximation to the seismograms in the unobserved low frequency band, both in phase and in amplitude. The numerical experiments considered are set up on simulated P-wave data. In extrapolated FWI (EFWI), the low-wavenumber components of the model are determined from the extrapolated low frequencies, before proceeding with a frequency sweep of the bandlimited data. The proposed deep-learning method of low-frequency extrapolation shows adequate generalizability for the initialization step of EFWI. Numerical examples show that the neural network trained on several submodels of the Marmousi model is able to predict the low frequencies for the BP 2004 benchmark model. Additionally, the neural network can robustly process seismic data with uncertainties due to the existence of noise, poorly-known source wavelet, and different finite-difference scheme in the forward modeling operator. Finally, this approach is not subject to strong assumptions of other methods for bandwidth extension, and seems to offer a tantalizing solution to the problem of properly initializing FWI.
\end{abstract}


\section{Acknowledgments}
The authors thank Total SA for support. LD is also supported by AFOSR grant FA9550-17-1-0316. Tensorflow and Keras are used for deep learning. The Python Seismic Inversion Toolbox (PySIT) \citep{PySIT2013} is used for FWI in this paper.

\section{Introduction}
FWI requires low frequency data to avoid convergence to a local minimum in the case where the initial models miss a reasonable representation of the complex structure. However, because of the acquisition limitation in seismic processing, the input data for seismic inversion are typically limited to a band above 3Hz. With assumptions and approximations to make inferences from tractable but simplified models, geophysicists have started reconstructing the reflectivity spectrum from the bandlimited records by signal processing methods. L$_1$-norm minimization \citep{levy1981reconstruction, oldenburg1983recovery}, autoregressive modelling \citep{walker1983autoregressive} and minimum entropy reconstruction \citep{sacchi1994minimum} have been developed to recover the isolated spikes of seismic recordings. Recently, bandwidth extension to the low frequency band has attracted the attention of many people in terms of FWI. For example, they recover the low frequencies by the envelope of the signal \citep{wu2014seismic, hu2017adaptive} or the inversion of the reflectivity series and convolution with the broadband source wavelet \citep{wang2016frequency, zhang2017sparse}. However, the low frequencies recovered by these methods are still far from the true low frequency data. \citet{li2016full} attempt to extrapolate the true low frequency data based on the phase tracking method \citep{li2015phase}. Unlike the explicit parameterization of phases and amplitudes of atomic events, here we propose an approach that can automatically process the raw bandlimited records. The deep neural network (DNN) is trained to automatically recover the missing low frequencies from the input bandlimited data. 

Because of the state-of-the-art performance of machine learning in many fields, geophysicists have begun adapting such ideas in seismic processing and interpretation \citep{chen2017automated, guitton2017statistical, xiong2018seismic}. By learning the probability of salt geobodies being present at any location in a seismic image, \citet{lewis2017deep} investigate CNN to incorporate the long wavelength features of the model in the regularization term. \citet{richardson2018seismic} constructs FWI as recurrent neural networks. \citet{araya2018deep, wu2018inversionnet,li2019deep} produce layered velocity models from shot gathers with DNN.

Like these authors and many others, we have selected DNN for low frequency extrapolation due to the increasing community agreement in favor of this method as a reasonable surrogate for a physics-based process \citep{grzeszczuk1998neuroanimator, de2011physics, araya2017automated}. The universal approximation theorem also indicates that the neural networks can be used to replicate any function up to our desired accuracy if the DNN has sufficient hidden layers and nodes \citep{hornik1989multilayer}. Although training is therefore expected to succeed arbitrarily well, only empirical evidence currently exists for the often-favorable performance of testing a network out of sample. Furthermore, we choose to focus on DNN with a convolutional structure, i.e., CNN. The idea behind CNN is to mine the hidden correlations among different frequency components.

In the case of bandwidth extension, the relevant data are the amplitudes and phases of seismic waves, which are dictated by the physics of wave propagation. For training, large volumes of synthetic shot gathers are generated from different models, in a wide band that includes the low frequencies, and the network's parameters are fit to regress the low frequencies of those data from the high frequencies. The window to split the spectrum to low and high frequency band should be smooth in frequency domain. For testing, bandlimited (and not otherwise processed) data from a new geophysical scenario are used as input of the network, and the network generates a prediction of the low frequencies. In the synthetic case, validation of the testing step is possible by computing those low frequencies directly from the wave solver.

By now, neural networks have shown their ability to fulfill the task of low frequency extrapolation. \citet{ovcharenko2017neural, ovcharenko2018low, ovcharenko2019deep, ovcharenko2019transfer} train neural network on data generated for random velocity models \citep[]{kazei2019realistically} to predict single low frequency from multiple high frequency data. They treat each shot gather in the frequency domain as a digital image for feature detection and thus require a large number of numerical simulations to synthesize the training data. \citet{jin2018learn} and \citet{hu2019progressive} use a deep inception based convolutional networks to synthesize data at multiple low frequencies. The input of their neural network contains the phase information of the true low frequency by leveraging the beat tone data \citep[]{hu2014fwi}. In contrast, we design an architecture of CNN to directly deal with the bandlimited data in the time domain. The proposed architecture can flexibly import one trace or multiple traces of the bandlimited shot gather to predict the data in a low frequency band with high enough accuracy that it can be used for FWI.

The limitations of neural networks for such signal processing tasks, however, are (1) the unreliability of the prediction when the training set is insufficient, and (2) the absence of a physical interpretation for the operations performed by the network. In addition, no theory can currently explain the generalizability of a deep network, i.e., the ability to perform nearly as well on testing as on training in a broad range of cases. Even so, the numerical examples indicate that the proposed architecture of CNN enjoys sufficient generalizability to extrapolate the low frequencies of unknown subsurface structures, in a range of numerical experiments.

We demonstrate the reliability of the extrapolated low frequencies to seed frequency-sweep FWI on the Marmousi model and the BP 2004 benchmark model. Two precautions are taken to ensure that trivial deconvolution of a noiseless record (by division by the high frequency (HF) wavelet in the frequency domain) is not an option: (1) add noise to the testing records, and (2) for testing, choose a hard bandpass HF wavelet taken to be zero in the low frequency (LF) band. In one numerical experiment involving bandlimited data above 0.6Hz from the BP 2004 model, the inversion results indicate that the predicted low frequencies are adequate to initialize conventional FWI from an uninformative initial model, so that it does not suffer from the otherwise-inherent cycle-skipping at 0.6Hz. Additionally, the proposed neural network has acceptable robustness to uncertainties due to the existence of noise, poorly-known source wavelet, and different finite-difference schemes in the forward modeling operator. 

This paper is organized as follows. We start by formulating bandwidth extension as a regression problem in machine learning. Next, we introduce the general workflow to predict the low frequency recordings with CNN. We then study the generalizability and the stability of the proposed architecture in more complex situations. Last, we illustrate the reliability of the extrapolated low frequencies to initialize FWI, and analyze the limitations of this method.

\section{Deep Learning}
A neural network defines a mapping $\textbf{y}=f(\textbf{x},\textbf{w})$ and learns the value of the parameters $\textbf{w}$ that result in a good fit between $\textbf{x}$ and $\textbf{y}$. DNNs are typically represented by composing together many different functions to find complex nonlinear relationships. The chain structures are the most common structures in DNNs \citep{goodfellow2016deep}:
\begin{equation}
\textbf{y} = f(\textbf{x},\textbf{w}) = f_L(...f_2(f_1(\textbf{x}))),
\end{equation}
where $f_1, f_2$ and $f_L$ are the first, the second and the $L^{th}$ layer of the network (with their own parameters omitted in this notation). Each $f_j$ consists of three operations taken in succession: an affine (linear plus constant) transformation, a batch normalization (multiplication by a scalar chosen adaptively), and the componentwise application of a nonlinear activation function. It is the nonlinearity of the activation function that enables the neural network to be a universal function approximator. The overall length $L$ of the chain gives the depth of the deep learning model. The final layer is the output layer, which defines the size and type of the output data. The training sets specify directly what the output layer must do at each point $\textbf{x}$, and constrain but do not specify the behavior of the other hidden layers. Rectified activation units are essential for the recent success of DNNs because they can accelerate convergence of the training procedure. Our numerical experiments show that, for bandwidth extension, Parametric Rectified Linear Unit (PReLU)\citep{he2015delving} works better than the Rectified Linear Unit (ReLU). The formula of PReLU is 
\begin{equation}  
g(\alpha, \textbf{y}) = 
\left\{  
\begin{array}{lr}  
\alpha\textbf{y}, & if \quad \textbf{y} < 0 \\  
\textbf{y}, \quad  & if \quad \textbf{y} \ge 0
\end{array}  ,
\right.  
\end{equation}  
where $\alpha$ is also a learned parameter and would be adaptively updated for each rectifier during training. 

Unlike the classification problem that trains the DNNs to produce discrete labels, the regression problem trains the DNNs for the prediction of continuous-valued outputs. It evaluates the performance of the model by means of the mean-squared error (MSE) of the predicted outputs $f(\textbf{x}_i;\textbf{w})$ vs. actual outputs $\textbf{y}_i$:
\begin{equation}
J (\textbf{w}) = \frac{1}{m} \sum_{i=1}^m L(\textbf{y}_i,f(\textbf{x}_i,\textbf{w})),
\end{equation}
where the loss $L$ is the squared error between the true low frequencies and the estimated outputs of the neural networks. The cost function $J$ is here minimized over $\textbf{w}$ by a stochastic gradient descent (SGD) algorithm, where each gradient is computed from a mini-batch, i.e., a subset in a disjoint randomized partition of the training set. Each gradient evaluation is called an iteration, while the full pass of the training algorithm over the entire training set using mini-batches is an epoch. The learning rate $\eta$ (step size) is a key parameter for deep learning and must be finetuned.
The gradients $\frac{\partial J(\textbf{w}^t)}{\partial\textbf{w}}$ of the neural networks are calculated by the backpropagation method \citep{goodfellow2016deep}.

CNN is an overwhelmingly popular architecture of DNN to extract spatial features in image processing, and it is the choice that we make in this paper. In this case, the matrix-vector multiplication in each of the $f_j$ is a convolution. In addition, imposing local connections and weight sharing can exploit both the local correlation and global features of the input image. CNNs are normally designed to deal with the image classification problem. For bandwidth extension, the data to be learned are the time-domain seismic signals, so we directly consider the amplitude at each sampling point as the pixel value of the image to be used as input of the CNN. 

Recall that CNN involve stacks of: a convolutional layer, followed by a PReLU layer, and a batch normalization layer. The filter number in each convolutional layer determines the dimensionality of the feature map or the channel of its output. Each output channel of the convolutional layer is obtained by convolving the channel of the previous layer with one filter, summing and adding a bias term. The batch normalization layer can speed up training of CNNs and reduce the sensitivity to network initialization by normalizing each input channel across a mini-batch. Although a pooling layer is typically used in the conventional architecture of CNNs, we leave it out because both the input and output signals have the same length, so since downsampling of feature maps is unhelpful for bandwidth extension in our experiments.

An essential hyperparameter for low frequency extrapolation with deep learning is the receptive field of a neuron. It is the local region of the input volume that affects the response of this neuron -- otherwise known as the domain of dependence. The spatial extent of this connectivity is related to the filter size. Unlike the small filter size commonly used in the image classification problem, we directly use a large filter in the convolutional layer to increase the receptive field of the CNN quickly with depth. The large filter size gives the neural network enough freedom to reconstruct the long-wavelength information

\begin{figure}[H]
\centering
\includegraphics[width=1\linewidth]{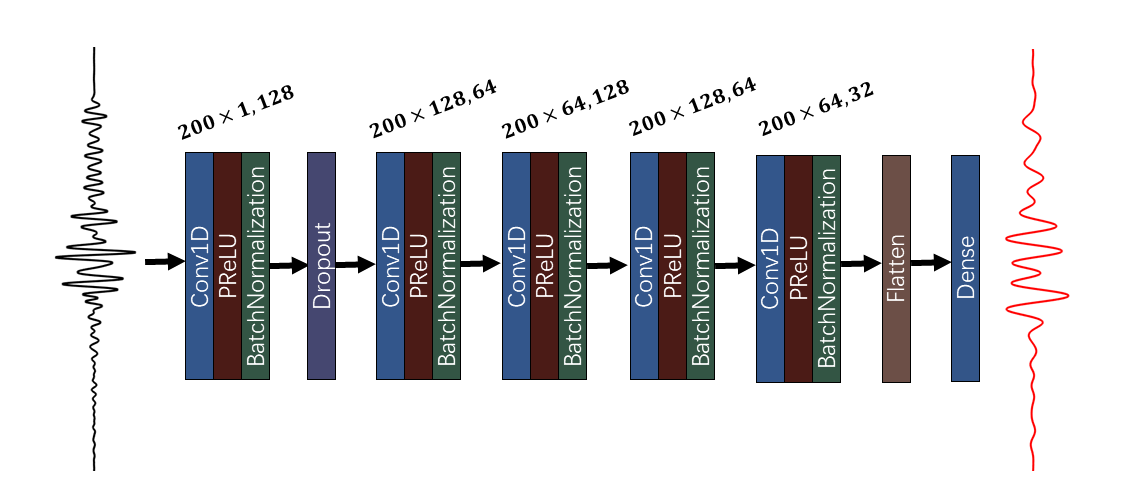}
\caption[]{An illustration of the architecture of our CNN to extrapolate the low frequency data from the bandlimited data in the time domain trace by trace. The architecture is a feed-forward stack of five sequential combinations of the convolution, PReLU and batch normalization layers, followed by one fully connected layer that outputs continuous-valued amplitude of the time-domain signal in the low frequency band. The network’s input is a one-dimentional bandlimited recording of length $nt$ where $nt$ is the number of time steps. The size and number of filters are labeled on the top of each convolutional layer.}
\label{fig:cnn}
\end{figure}

The architecture of our neural network (Figure~\ref{fig:cnn}) is a feed-forward stack of five sequential combinations of the convolution, PReLU and batch normalization layers, followed by one fully connected layer that outputs continuous-valued amplitude of the time-domain signal in the low frequency band. The first convolutional layer filters the $nt\times1$ input time series with 128 kernels of size $200\times1\times1$ where $nt$ is the number of time steps. The second convolutional layer has  64 kernels of size $200\times1\times128$ connected to the normalized outputs of the first convolutional layer. The third convolutional layer has 128 kernels of size $200\times1\times64$. The fourth convolutional layer has 64 kernels of size $200\times1\times128$ , and the fifth convolutional layer has 32 kernels of size $200\times1\times64$. The last layer is fully connected, taking features from the last convolutional layer as input in a vector form of length $nt\times32$. The stride of the convolution is one, and zero-padding is used to make the output length of each convolution layer the same as its input. Additionally, a dropout layer \citep{srivastava2014dropout} with a probability of $50\%$ is added after the first convolution layer to reduce the generalization error.

We use CNN in the context of supervised learning, i.e., inference of $\textbf{y}_i$ from $\textbf{x}_i$. We need to first train the CNN from a large number of samples $(\textbf{x}_i, \textbf{y}_i)$ to determine the coefficients of the network, and then use the network for testing on new $\textbf{x}_i$. In statistical learning theory, the generalization error is the difference between the expected and empirical error, where the expectation runs over a continuous probability distribution on the $\textbf{x}_i$. This generalization error can be approximated by the difference between the errors on the training and on the test sets. 

The object of this paper is that the $\textbf{x}_i$ can be taken to be seismograms bandlimited to the high frequencies, and $\textbf{y}_i$ can be the same seismograms in the low frequency band. Generating training samples means collecting, or synthesizing seismogram data from a variety of geophysical models, which enter as space-varying elastic coefficients in a wave equation. For the purpose of good generalization (small generalization error), the models used to create the large training sets should be able to represent many subsurface structures, including different types of reflectors and diffractors, so we can find a representative set of parameters to handle data from different scenarios or regions. The performance of the neural network is sensitive to the architecture and the hyperparameters, so we must design them carefully. Next, we illustrate the specific choice of hyperparameters for bandwidth extension, along with numerical examples involving synthetic data from community models.

\section{Numerical Examples}
In this section, we demonstrate the reliability of extrapolated FWI with CNN (EFWI-CNN) in three parts. In the first part, we show CNN's ability to extrapolate low frequency data ($0.1-5 $Hz) from bandlimited data ($5-20$Hz) on the Marmousi model (Figure~\ref{fig:training_model_position}). In the second part, we verify the robustness of the method with uncertainties in the seismic data due to the existence of noise, different finite difference scheme, and poorly-known source wavelet. In the last part, we perform EFWI-CNN on both the Marmousi model and the BP 2004 benchmark model \citep{billette20052004}, by firstly using the extrapolated low frequencies to synthesize the low-wavenumber background velocity model. Then, we compare the inversion results with the bandlimited data in three cases which respectively start FWI from an uninformed initial model, the low-wavenumber background model created from the extrapolated low frequencies, and the low-wavenumber background model created from the true low frequencies.

\subsection{Low frequency extrapolation}
\begin{figure}[H]
\centering
\includegraphics[width=\columnwidth]{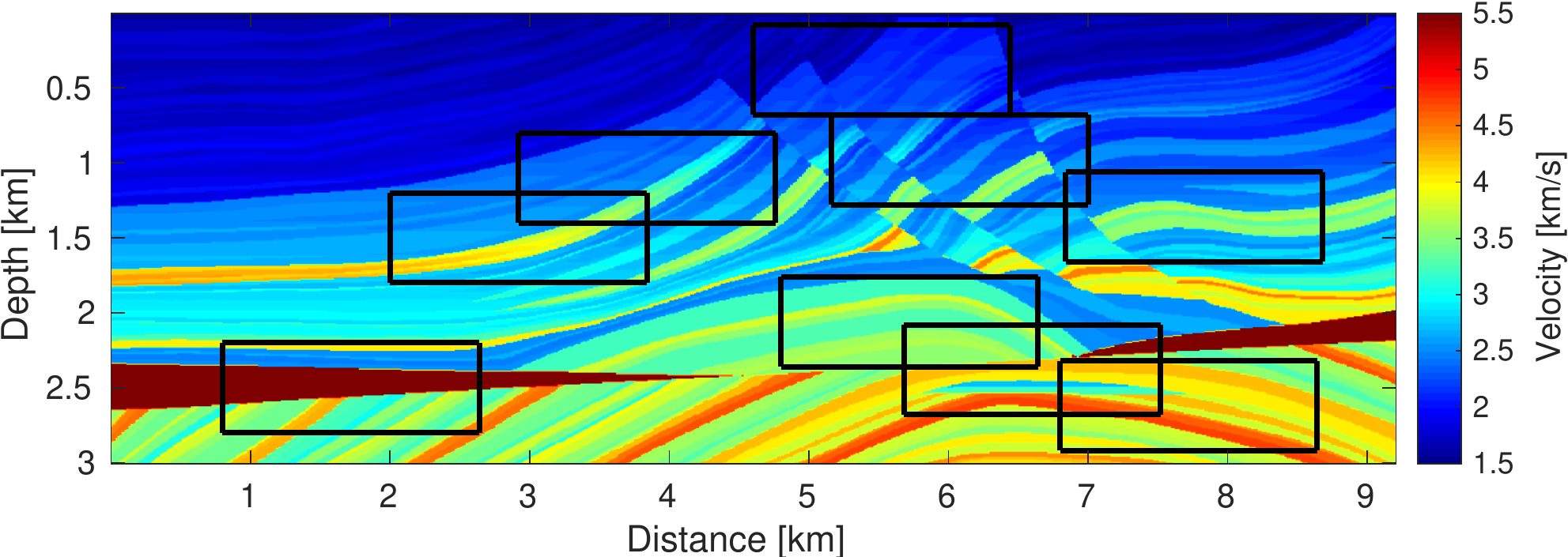}      
\caption{The nine training models randomly extracted from the Marmousi velocity model to collect the training data set. The test models are the Marmousi model and the BP 2004 benchmark model. A water layer with 300m depth is added to the top of these training models and Marmousi model. We use the same training models to extrapolate the low frequencies on both test models.}
\label{fig:training_model_position}
\end{figure}

Following our previous work \citep[]{sun2018low, sun2019extrapolated}, the true unknown velocity model for FWI is referred to as the test model, since it is used to collect the test data set in deep learning. To collect the training data set, we create training models by randomly selecting nine parts of the Marmousi model (Figure~\ref{fig:training_model_position}) with different structure but the same number of grid points $166\times461$. We also downsample the original model to $166\times461$ pixels as the test model. We find that the randomized models produced in this manner are realistic enough to demonstrate the generalization of the neural network if the structures of the submodels are diversified enough. 

\begin{figure}[H]
\centering
\includegraphics[width=\columnwidth]{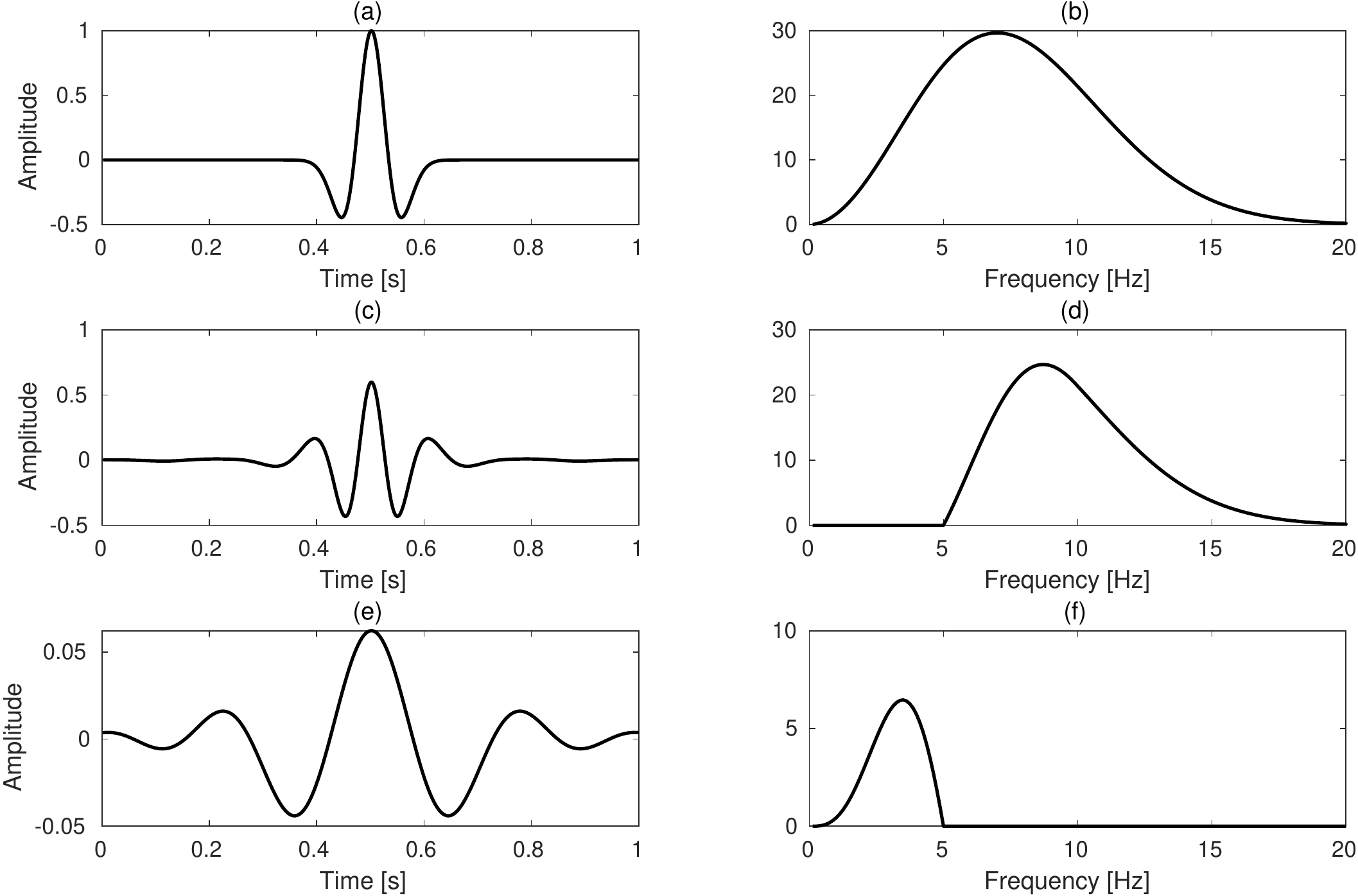}      
\caption{(a) The Ricker wavelet with 7Hz dominant and its amplitude spectrum in (b). (c) The high frequency wavelet bandpassed from (a) and its amplitude spectrum in (d). (e) The low frequency wavelet bandpassed from (a) and its amplitude spectrum in (f).}
\label{fig:compare_wavelet}
\end{figure}

In this example, we have the following processing steps to collect each sample (i.e., shot record) in both the training and test data sets.
\begin{itemize}  
\item The acquisition geometry of forward modeling on each model is the same. It consists of 30 sources and 461 receivers evenly spaced at the surface. We consider each time series or trace as one sample in the data set, so have $124,470$ training samples and $13,830$ test samples for each test model in total.
\item We use a fourth order in space and second order in time finite-difference modeling method with PML to solve the 2D acoustic wave equation in the time domain, to generate the synthetic shot gathers of both the training and test data sets. The sampling interval and the total recording time are 2ms and 5s, respectively.  
\item We use a Ricker wavelet with dominant frequency 7Hz to synthesize the full-band seismic recordings. Then the data below 5Hz and above 5Hz are split to synthesize the output and input of the neural networks, respectively (Figure~\ref{fig:compare_wavelet}). Both the low and high frequency data are obtained by a sharp windowing of the same trace.
\end{itemize}

\begin{figure}[H]
\centering
\includegraphics[width=0.5\columnwidth]{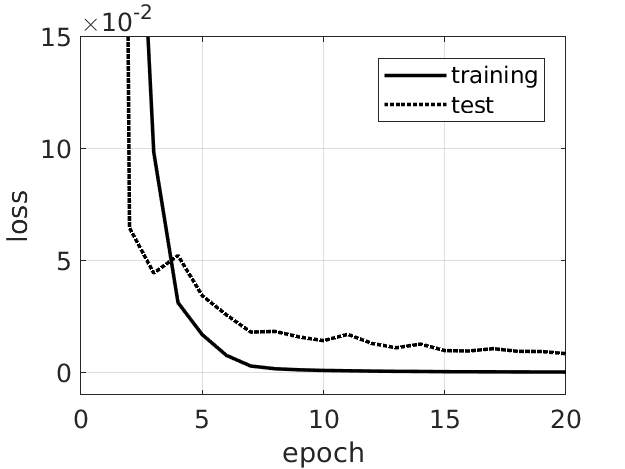}
\caption{The learning curves. Both training and test losses decay with the training steps.}
\label{fig:mar_training_loss}
\end{figure}

\begin{figure}[H]
\centering
\includegraphics[width=\columnwidth]{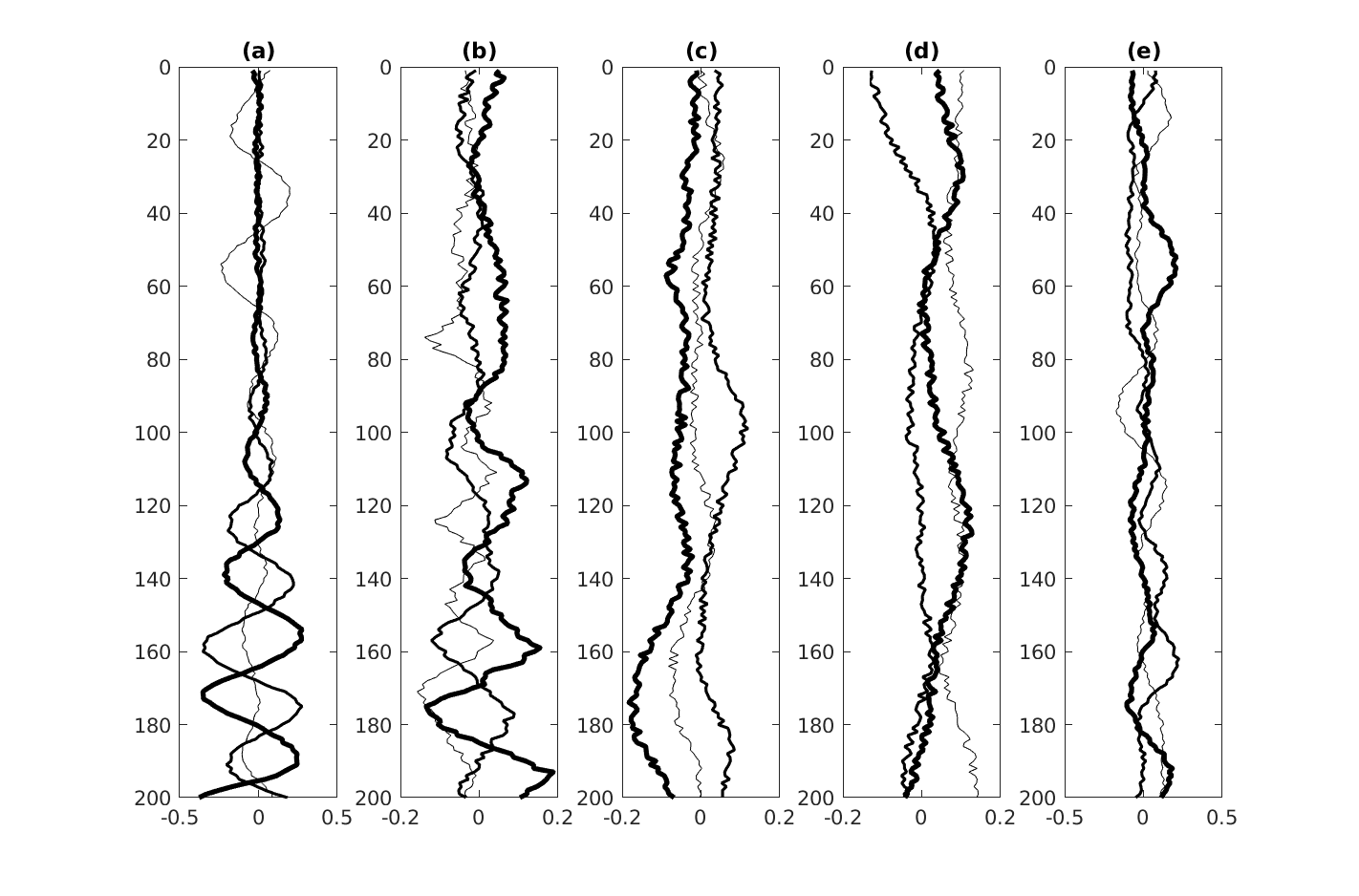}      
\caption{Three kernels of the first channel in the (a) first (b) second (c) third (d) fourth and (e) last convolutional layer learned by training with 20 epochs to predict the low frequency data below 5Hz on the Marmousi model.}
\label{fig:compare_filter_line}
\end{figure}

In this example, we train the network with the Adam optimizer and use a mini-batch of 64 samples at each iteration. The initial learning rate and forgetting rate of the Adam are the same as the original paper \citep{kingma2014adam}. The initial value of the bias is zero. The weight initialization is via the Glorot uniform initializer \citep{glorot2010understanding}. It randomly initializes the weights from a truncated normal distribution centered on zero with the standard deviation $\sqrt{2/{n_1+n_2}}$ where $n_1$ are $n_2$ are the numbers of input and output units in the weight tensor, respectively.

\begin{figure}[H]
\centering
\includegraphics[width=\columnwidth]{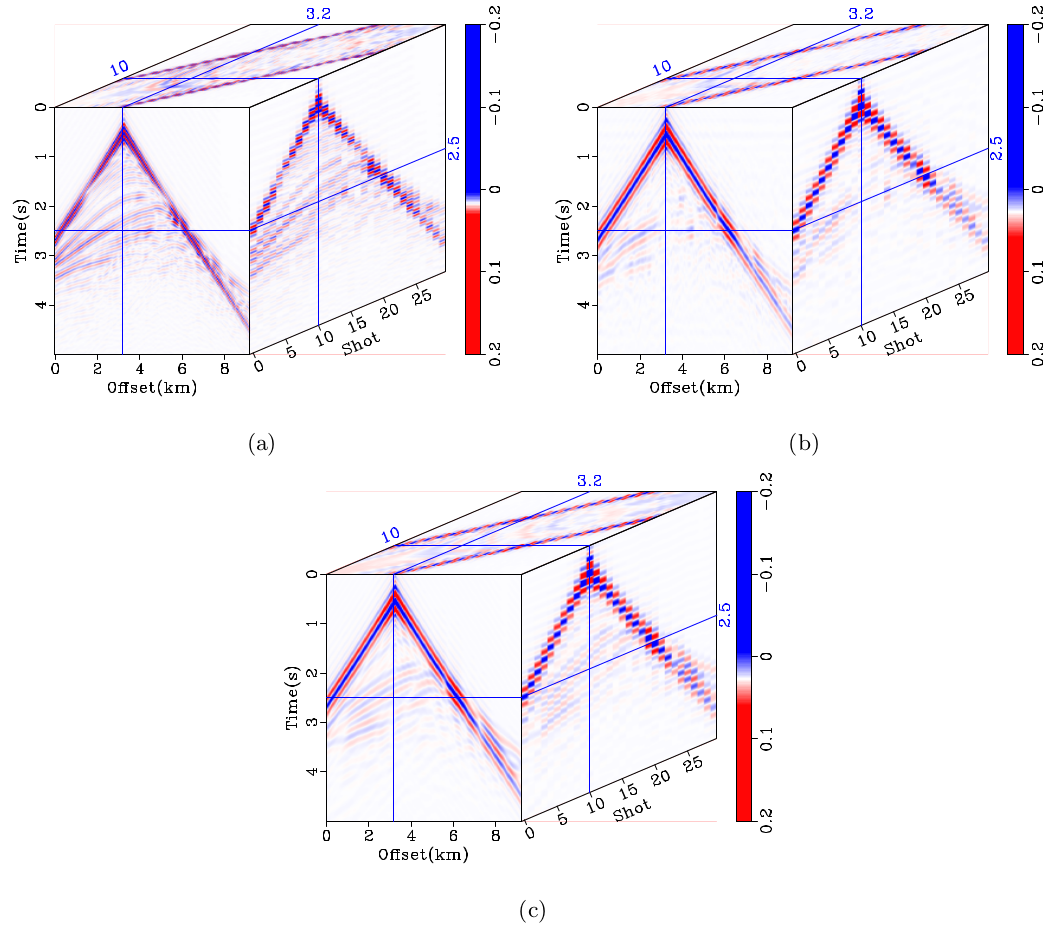}      
\caption{The extrapolation result on the Marmousi model: comparison among the (a) bandlimited recordings ($5-20$Hz), (b) predicted and (c) true low frequency recordings ($0.1-5$Hz). The bandlimited data in (a) are the inputs of CNNs to predict the low frequencies in (b).}
\label{fig:Figure6}
\end{figure}

\begin{figure}[H]
\centering
\includegraphics[width=1.0\columnwidth]{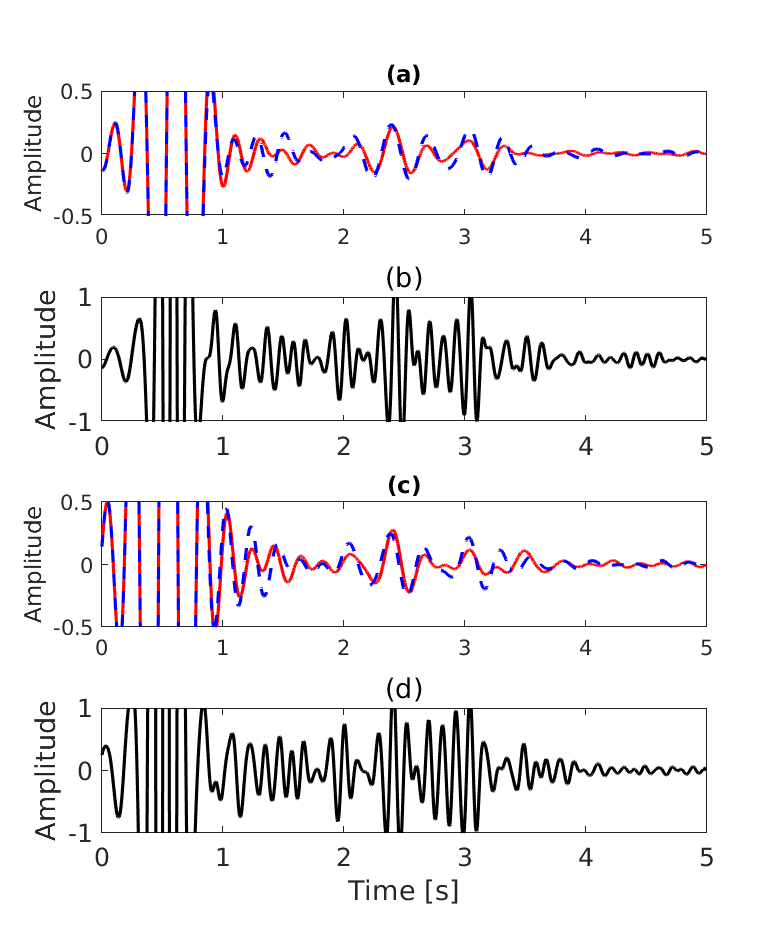}      
\caption{The extrapolation result on the Marmousi model: comparison among the predicted (red line), the true (blue dash line) recording in the low frequency band ($0.1-5$Hz) and the bandlimited recording (black line) ($5-20$Hz) at the horizontal distance (a) (b) $x=2.82km$ and (c) (d) $x=2.92km$.}
\label{fig:records_line_ishot_10_imodel_1_paper}
\end{figure}

\begin{figure}[H]
\centering
\includegraphics[width=1.0\columnwidth]{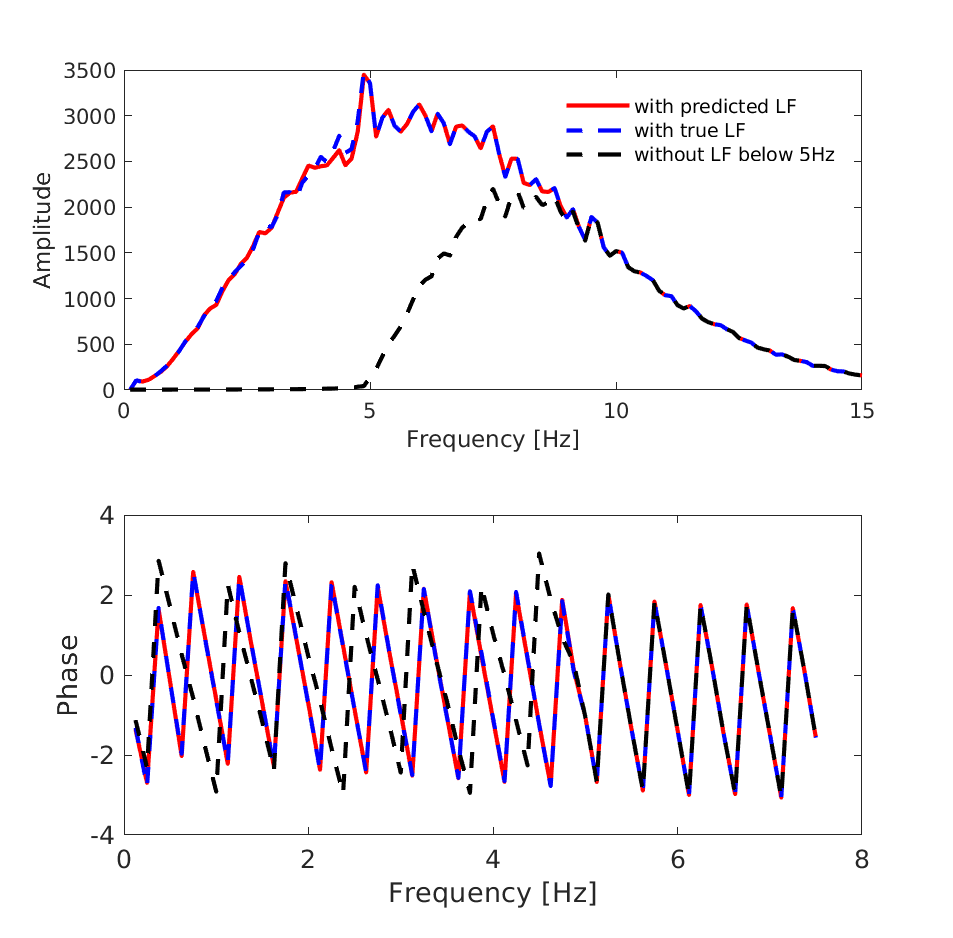}      
\caption{The extrapolation result on the Marmousi model: comparison of (a) the amplitude spectrum and (b) the phase spectrum at $x=2.92km$ among the bandlimited recording ($5-20$Hz), the recording ($0.1-20$Hz) with true and predicted low frequencies ($0.1-5$Hz).}
\label{fig:spectrum_line_ishot_10_itrace_147}
\end{figure}

The training process of the 20 epochs is shown in Figure~\ref{fig:mar_training_loss}. Both training and test losses decay with the training steps, which indicates that our neural network is not overfitting. Figure~\ref{fig:compare_filter_line} show the kernals of CNN learned by training with 20 epochs. Three representative filters of the first channel are plotted in each of the convolutional layers. However, it is still difficult to fully interpret the features. We test the performance of the neural networks by feeding the bandlimited data in the test set into the pretrained neural networks and obtain the extrapolated low frequencies on the Marmousi model. Figure~\ref{fig:Figure6} compares the shot gather between the bandlimited data ($5-20$Hz), extrapolated and true low frequencies ($0.1-5$Hz) where the source is located at the horizontal distance $x=2.94km$ on the Marmousi model. The extrapolated results in Figure~\ref{fig:Figure6}(c) show that the proposed neural network can accurately predict the recordings in the low frequency band, which are totally missing before the test. Figure~\ref{fig:records_line_ishot_10_imodel_1_paper} compares two individual seismograms in Figure~\ref{fig:Figure6}(c) where the receivers are located at the horizontal distance $x=2.82km$ and $x=2.92km$, respectively. The extrapolated low frequency data match the true recordings well. Then we combine the extrapolated low frequencies with the bandlimited data and compare the amplitude spectrum of the full-band data with the extrapolated and true low frequencies. The amplitude and phase spectrum comparison of the single trace where the receiver is located at $x=2.92km$ (Figure \ref{fig:spectrum_line_ishot_10_itrace_147}) clearly shows that the neural networks can capture the relationship between low and high frequency components constrained by the wave equation.

\begin{figure}[H]
\centering
\includegraphics[width=\columnwidth]{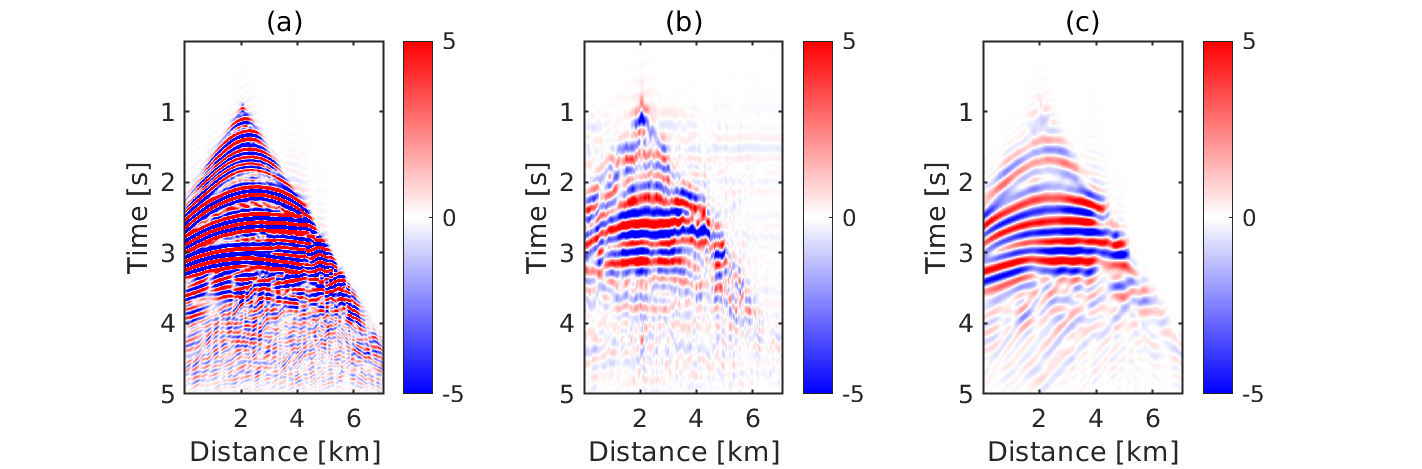}      
\caption{Low frequency extrapolation without direct waves: comparison among the (a) bandlimited recordings ($5-20$Hz), (b) predicted and (c) true low frequency recordings ($0.1-5$Hz) on the Marmousi model. The direct waves are muted from the full-band shot gathers with a smooth time window before spliting into the bandlimited recordings and the low frequencies. The extrapolation is robust with the presence of muting.}
\label{fig:records_ishot_7_imodel_1}
\end{figure}

Figure~\ref{fig:records_ishot_7_imodel_1} shows the low frequency extrapolation without direct waves. The direct waves are muted from the full-band shot gathers with a smooth time window before spliting into the bandlimited recordings and the low frequencies. The low frequencies of reflections are recovered without the existence of the direct waves. Therefore, the neural network is robust with the presence of muting.

\subsection{Uncertainty analysis}
With a view toward dealing with complex field data, we investigate the stability of the neural network's predictive performance under three kinds of discrepancies, or uncertainties, between training and test: additive noise; different finite difference operator in the forward modeling; and different source wavelet. In every case, we compare the extrapolation accuracy with the reference in Figure~\ref{fig:Figure6}, where training and testing are set up the same way (noiseless bandlimited data, finite difference operator with second order in time and fourth order in space, and the Ricker wavelet with 7Hz dominant frequency). The RMSE between the extrapolated and true low frequency data of the 30 shot gathers in Figure~\ref{fig:Figure6} is $2.1304\times10^{-4}$.

\begin{figure}[H]
\centering
\includegraphics[width=\columnwidth]{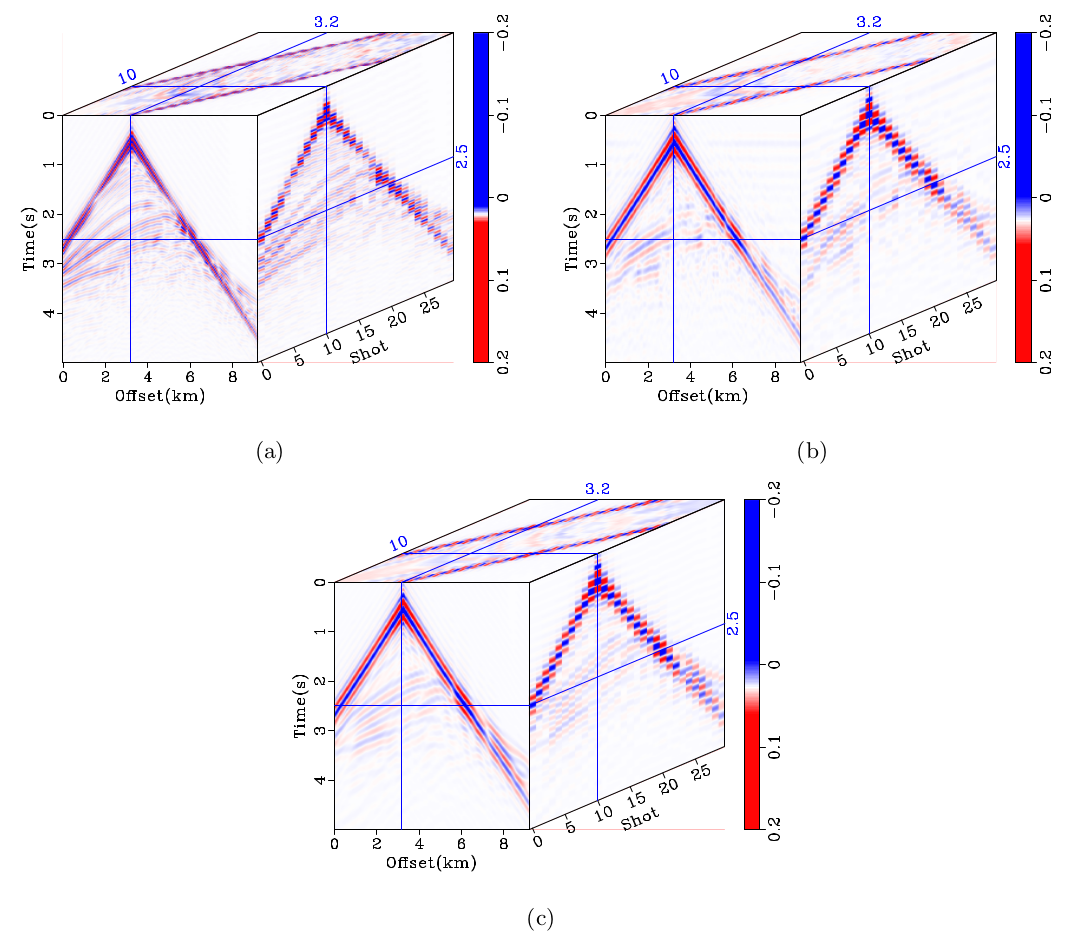}      
\caption{Noise robustness: comparison among the (a) bandlimited recordings ($5-20$Hz), (b) predicted and (c) true low frequency recordings ($0.1-5$Hz) on the Marmousi model. We add $25\%$ additive Gaussian noise to the bandlimited data in the test data set and $30\%$ additive Gaussian noise to the bandlimited data in the training data set. Even though noise will disturb the neural network find the correct mapping between the bandlimited data with their low frequencies, the proposed neural network can still extrapolate the low frequencies of the main reflections. The neural network is able to perform extrapolation as well as denoising.}
\label{fig:Figure10}
\end{figure}

In the first case, the neural network is expected to extrapolate the low frequencies from the noisy bandlimited data. We add additive $20\%$ Gaussian noise to the bandlimited data in the test data set and $30\%$ additive Gaussian noise to the bandlimited data in the training data set. The low frequencies in the training set are noiseless as before. Even though noise will disturb the neural network to find the correct mapping between the bandlimited data with their low frequencies, Figure~\ref{fig:Figure10} shows that the proposed neural network can still successfully extrapolate the low frequencies of the main reflections. The RMSE between the extrapolated and true low frequency data of the 30 shot gathers in Figure~\ref{fig:Figure10} is $2.4156\times10^{-4}$. The neural network is able to perform extrapolation as well as denoising. Incidentally, we make the (unsurprising) observation that CNN has potential for the denoising of seismic data.

\begin{figure}[H]
\centering
\includegraphics[width=\columnwidth]{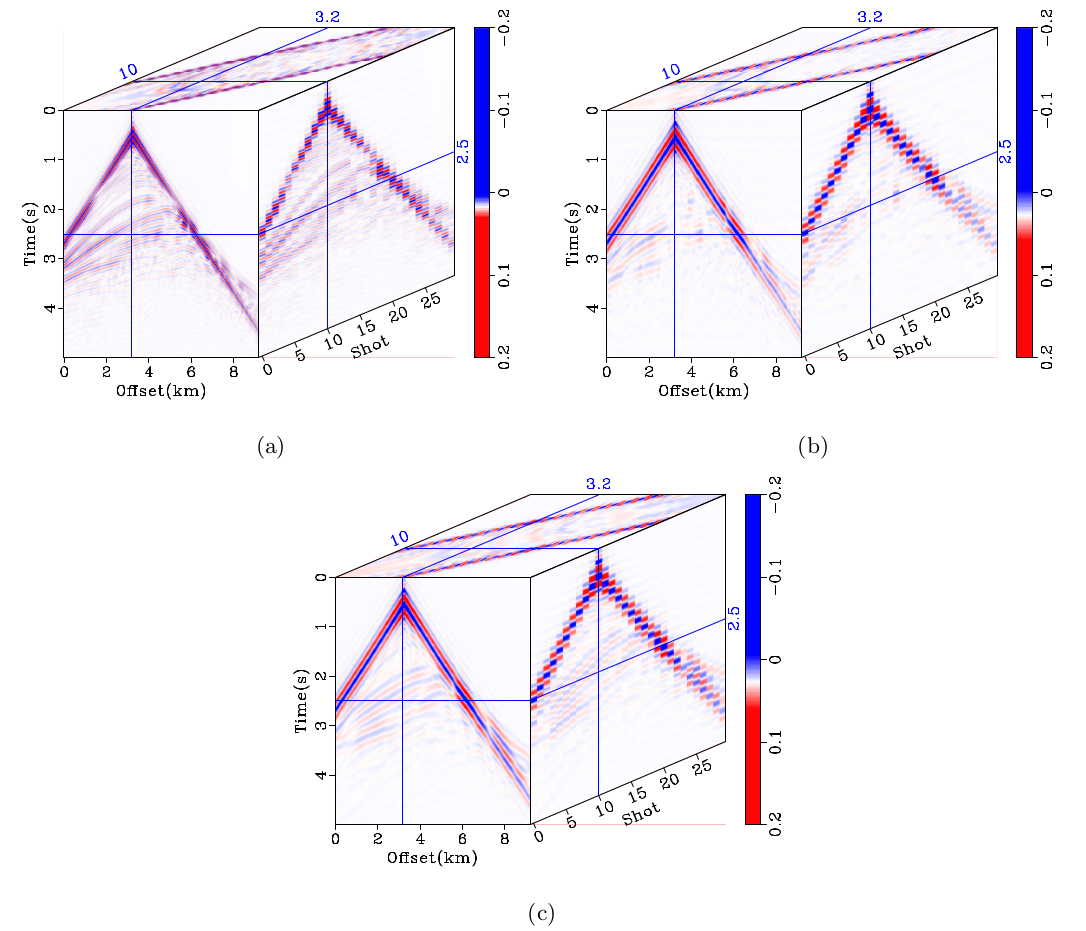}      
\caption{Forward modeling operator robustness: comparison among the (a) bandlimited recordings ($5-20$Hz), (b) predicted and (c) true low frequency recordings ($0.1-5$Hz) on the Marmousi model. The shot gather in the test data set is simulated with the sixth order operator, while the neural network is trained with the samples simulated with the fourth order operator. The extrapolation result in (b) shows that the neural network trained on the fourth order finite difference operator can extrapolate the low frequencies of the bandlimited data coming from the sixth order operator. The neural network is stable with the variance of the forward modeling operator.}
\label{fig:Figure11}
\end{figure}

Another challenge of FWI is that the observed and calculated data can come from different wave propagation schemes. For example, under the control of different numerical dispersion curves, the phase velocity would have different behaviors if we used different finite difference (FD) operators to simulate the observed and calculated data. Therefore, it is necessary to study the influence of different discretization, or other details of the simulation, on the accuracy of low frequency extrapolation. In our case, the shot gathers in the test data set are simulated with a sixth order spatial FD operator, but the neural network is trained on the samples simulated with a fourth order spatial FD operator. The extrapolation result in Figure~\ref{fig:Figure11}(b) shows that the neural network trained on the fourth order operator is able to extrapolate the low frequencies of the bandlimited data collected with the sixth order operator. In this case, the RMSE between the extrapolated (Figure~\ref{fig:Figure11}b) and true (Figure~\ref{fig:Figure11}c) low frequency data of the 30 shot gathers is $2.2248\times10^{-4}$. The neural network appears to be stable with respect to mild modifications to the forward modeling operator, at least in the examples tried.

\begin{figure}[H]
\centering
\includegraphics[width=\columnwidth]{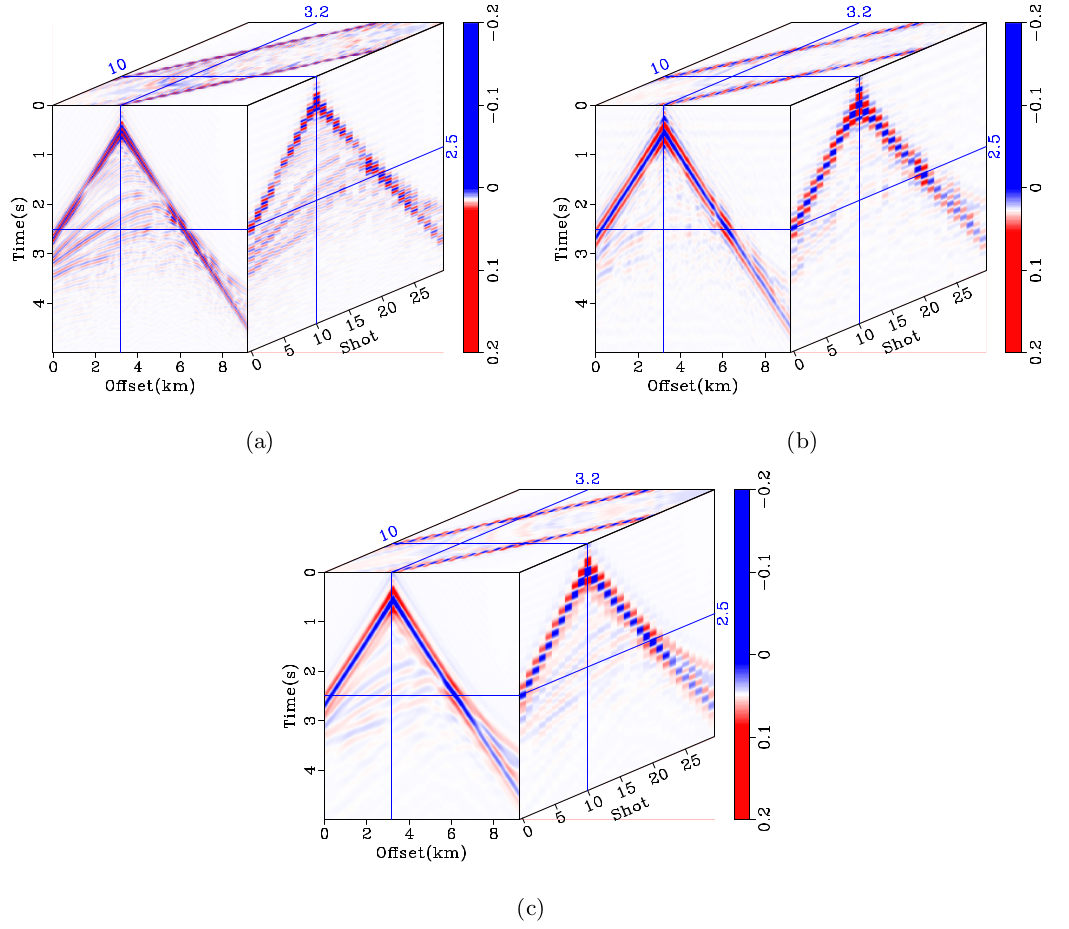}      
\caption{Unknown source wavelet robustness: comparison among the (a) bandlimited recordings ($5-20$Hz), (b) predicted and (c) true low frequency recordings ($0.1-5$Hz) on the Marmousi model. In this case, we use an Ormsby wavelet with the four corner frequencies 0.2Hz, 1.5Hz, 8Hz and 14Hz to synthesize the output and input of neural network for samples in the test data set. The result in (b) shows that the neural network trained on the data from 7Hz Ricker wavelet is able to extrapolate the data synthesize with an Ormsby wavelet. However, the recover of the amplitude is poorer than the phase.}
\label{fig:Figure12}
\end{figure}

Another uncertainty is the unknown source wavelet. To check the extrapolation capability of the neural network in the context of data excited by an unknown source wavelet, we train the neural network with a 7Hz Ricker wavelet, but test it with an Ormsby wavelet. The four corner frequencies of the Ormsby wavelet are 0.2Hz, 1.5Hz, 8Hz and 14Hz, respectively. Figure~\ref{fig:Figure12} shows that the neural network trained on the data from the 7Hz Ricker wavelet source wavelet is able to extrapolate the data synthesized with Ormsby source wavelet. However, the recover of the amplitude is much poorer than the phase. The RMSE between the extrapolated and true low frequency data of the 30 shot gathers in Figure~\ref{fig:Figure12} is increased to $1.1717\times10^{-3}$. The commonplace issue of the source wavelet being unknown or poorly known in FWI has seemingly little effect on the performance of the proposed neural network to extrapolate the phase of low frequency data, at least in the examples tried.

Even though all of the uncertain factors hurt the accuracy of extrapolated low frequencies to some extent, the CNN's prediction has a degree of robustness that surprised us. All of the extrapolation results in the above numerical examples can be further improved by increasing the diversity of the training data set, because subjecting the network to a broader range of scenarios can fundamentally reduce the generalization error of the deep learning predictor. (For example, we can simulate the training data set with multiple kinds of source wavelets and finite difference operators) 

\subsection{Extrapolated FWI: Marmousi model}
\begin{figure}[H]
\centering
\includegraphics[width=\columnwidth]{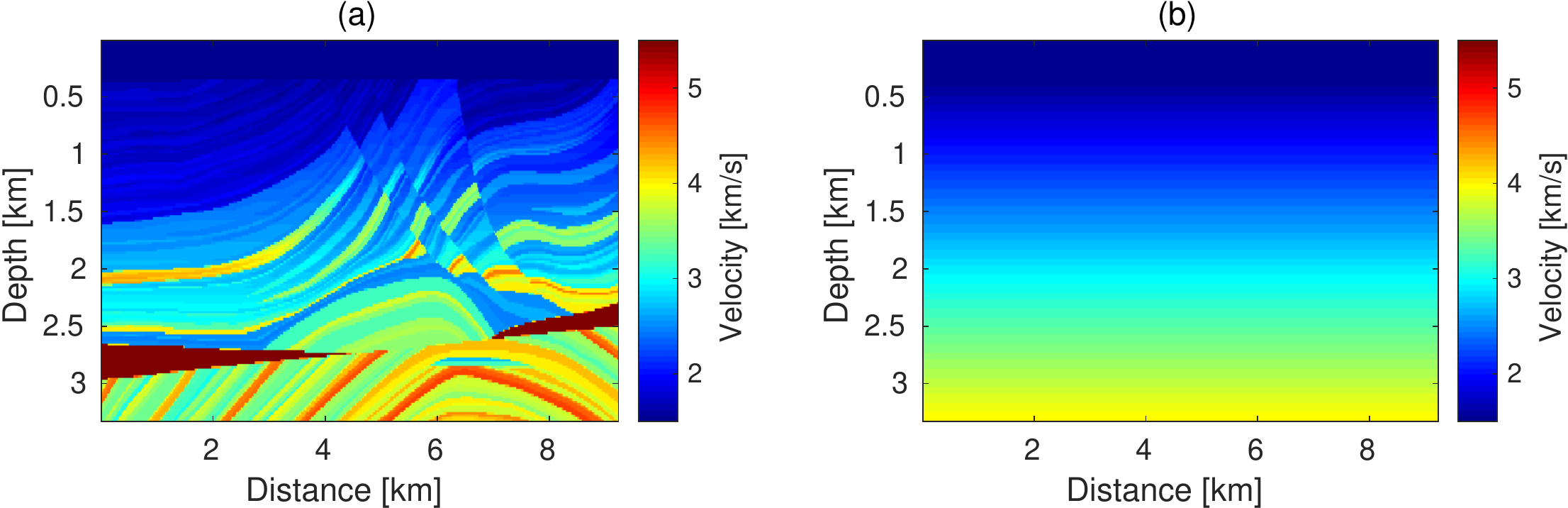}      
\caption{(a) The Marmousi model (the true model in FWI and the test model in deep learning) and (b) the initial model for FWI.}
\label{fig:compare_vtrue_and_v0}
\end{figure}

In this example, we construct the low-wavenumber velocity model for the Marmousi model, by leveraging the extrapolated low frequency data (Figure~\ref{fig:Figure6}(b)) to solve the cycle-skipping problem for FWI on the bandlimited data. The objective function of the inversion is formulated as the least-squares misfit between the observed and calculated data in the time domain. Starting from an initial model in Figure~\ref{fig:compare_vtrue_and_v0}(b), we use the LBFGS \citep{nocedal2006numerical} optimization method to update the model gradually. To help the gradient-based iterative inversion method avoid falling into local minima, we also perform the inversion from the lowest frequency that the data contained, to successively higher frequencies.

\begin{figure}[H]
\centering
\includegraphics[width=\columnwidth]{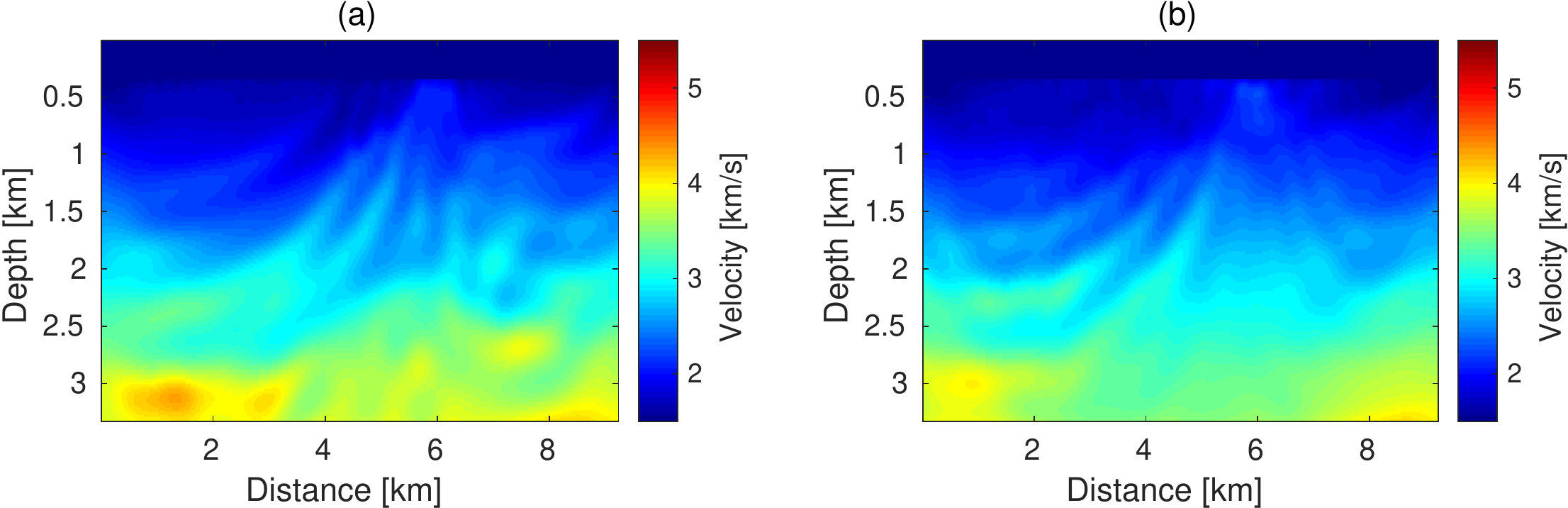}      
\caption{Comparison between the inverted low-wavenumber models using the (a) true and (b) extrapolated 0.5-3Hz low frequencies. The model constructed from the extrapolated low frequencies has lower resolution in the deeper section compared with the model from the true low frequencies because the extrapolation accuracy of deeper reflections is poor. However, both models capture the low wavenumber information of the Marmousi model.}
\label{fig:compare_initial}
\end{figure}

With this inversion scheme, we test the reliability of the extrapolated low frequencies (Figure~\ref{fig:Figure6}(b)) on the Marmousi model (Figure~\ref{fig:compare_vtrue_and_v0}(a)). The velocity structure of the initial model is far from the true model. The true model was not used in the training stage. The acquisition geometry and source wavelet are the same as in the example in the previous section. The observed data below 5.0Hz are totally missing. Therefore, we firstly use the bandlimited data in $5-20$Hz to recover the low frequencies in ($0.1-5.0$Hz) and then use the low frequencies to invert the low-wavenumber velocity model for the bandlimited FWI. Figure~\ref{fig:compare_initial} compares the inverted models from FWI using the true and extrapolated $0.5-3$Hz low frequency data. Since the low frequency extrapolation accuracy of reflections after four seconds is limited (as seen in Figure~\ref{fig:Figure6}), the low-wavenumber model constructed from the extrapolated low frequencies has lower resolution in the deeper section compared with that from the true low frequencies. However, both models capture the low wavenumber information of the Marmousi model. These models are used as the starting models for FWI on the bandlimited data.

\begin{figure}[H]
\centering
\includegraphics[width=\columnwidth]{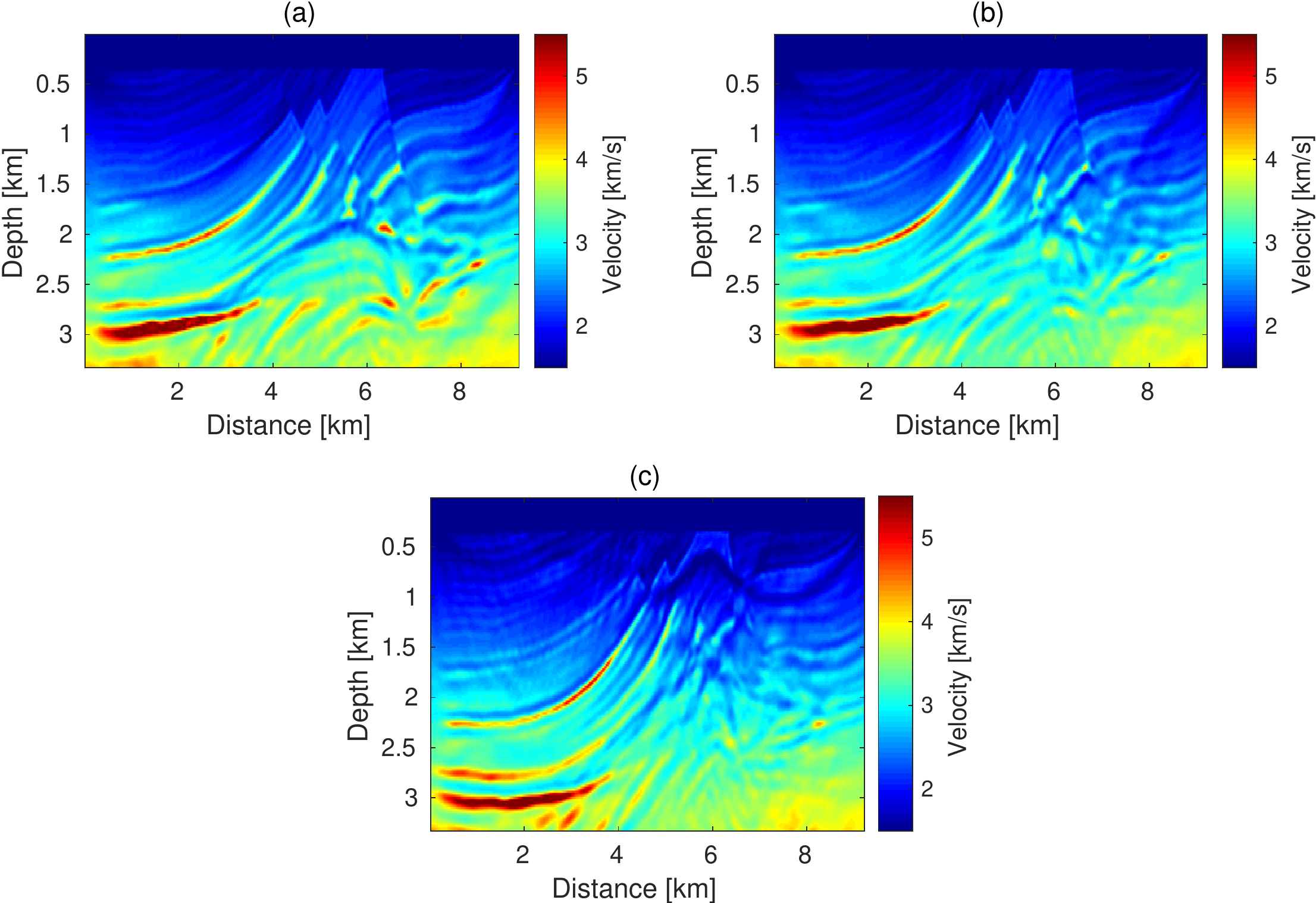}      
\caption{Comparison among the inverted models from FWI using the bandlimited data (5-15Hz). In (a), resulting model is started from the low-wavenumber velocity model constructed with the true low frequencies in Figure~\ref{fig:compare_initial}(a). In (b), resulting model is started from the low-wavenumber velocity model constructed with the extrapolated low frequencies in Figure~\ref{fig:compare_initial}(b). In (c), resulting model is started from the initial model in Figure~\ref{fig:compare_vtrue_and_v0}(b).}
\label{fig:compare_inversion}
\end{figure}

Figure~\ref{fig:compare_inversion} compares the inverted models from FWI using the bandlimited data (5-15Hz) with different starting models. The resulting model in Figure~\ref{fig:compare_inversion}(b) starts from the low-wavenumber model constructed from the extrapolated low frequencies (Figure~\ref{fig:compare_initial}(b)), which is almost the same as the one from the true low frequencies (Figure~\ref{fig:compare_inversion}(a)). Since the highest frequency component in the low frequency band is 3Hz when we invert the starting model, both inversion results have slight cycle skipping phenomenon. However, Figure~\ref{fig:compare_inversion}(c) performs the bandlimited FWI with the linear initial model, and shows a much more pronounced effect of cycle skipping. We cannot find much meaningful information about the subsurface structure if the bandlimited inversion starts at 5Hz from a linear initial model (Figure~\ref{fig:compare_vtrue_and_v0}(b)). 

\begin{figure}[H]
\centering
\includegraphics[width=\columnwidth]{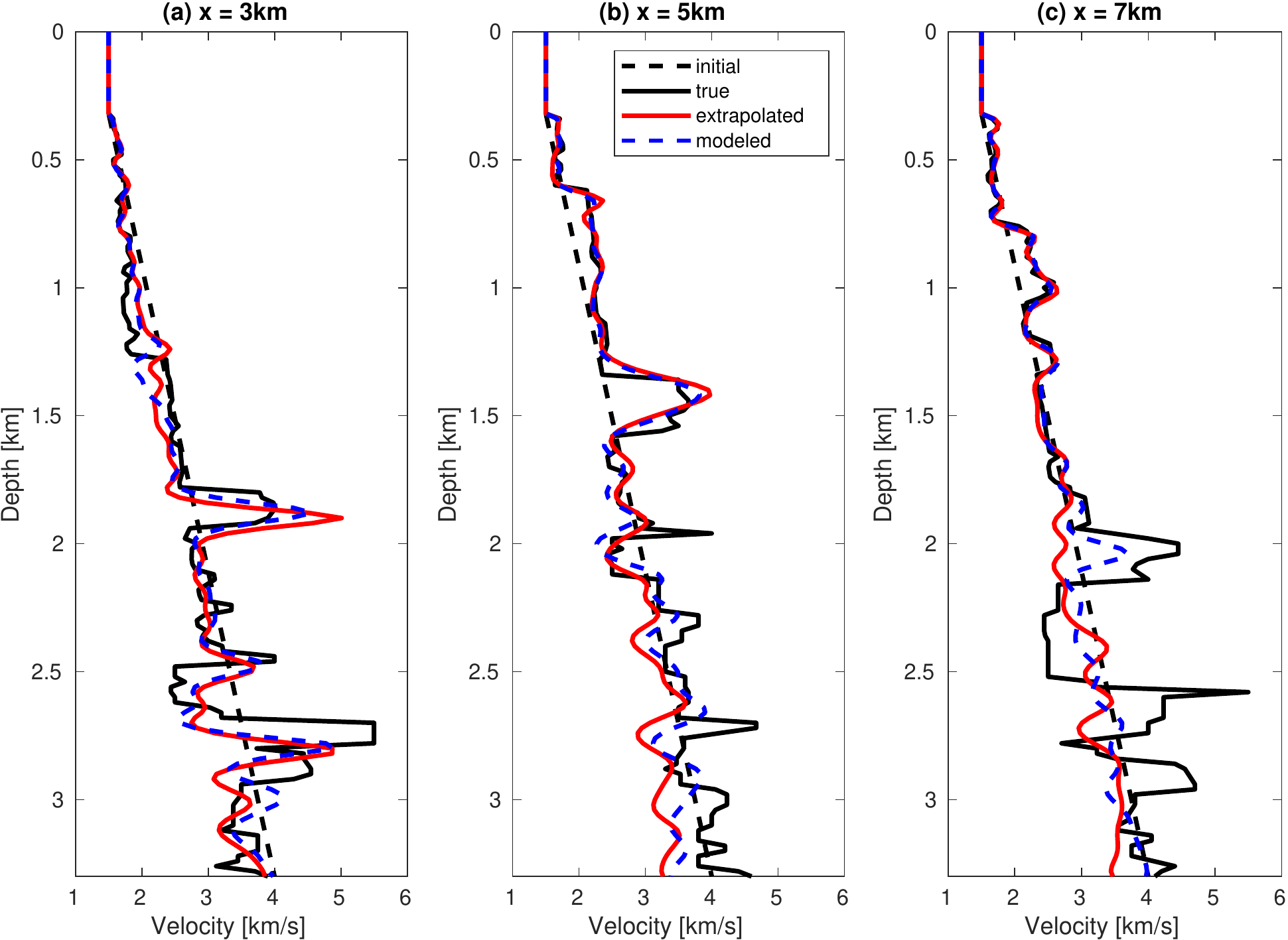}      
\caption{Comparison of velocity profiles among initial model (black dash line), true model (black line) and the resulting models started from the low-wavenumber models constructed with extrapolated (red line) and true (blue dash line) low frequencies at the horizontal locations of (a) x=3km, (b) x=5km and (c) x=7km.}
\label{fig:compare_inversion_line}
\end{figure}

Figure~\ref{fig:compare_inversion_line} compares the velocity profile among the resulting models in Figure~\ref{fig:compare_inversion} (the initial and true velocity models) at the horizontal location of $x=3km$, $x=5km$ and $x=7km$. The final inversion result started from the extrapolated low frequencies gives us almost the same model as the true low frequencies, which illustrates that the extrapolated low frequency data are reliable enough to provide an adequate low-wavenumber velocity model. However, both inversion workflows have difficulty in the recovery of velocity structure below 2km. The inversion results can be further improved by involving higher frequency components and adding more iterations. In contrast, since the velocity model in Figure~\ref{fig:compare_inversion}(c) has fallen into a local minimum, the inversion cannot converge to the true model in the subsequent iterations.

\subsection{Extrapolated FWI: BP model}
\begin{figure}[H]
\centering
\includegraphics[width=\columnwidth]{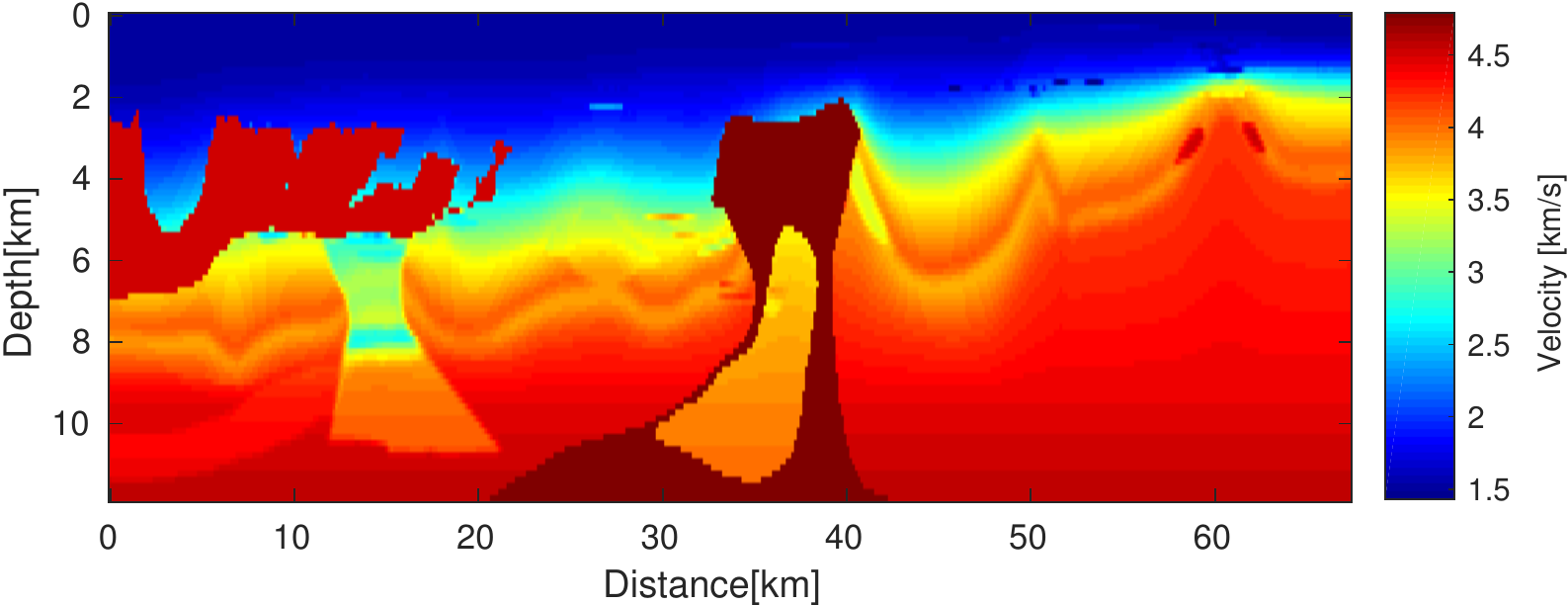}      
\caption{The 2004 BP benchmark velocity model used to collect the test data set for studying the generalizability of the proposed neural network. This model is the true model in extrapolated FWI.}
\label{fig:bp_vtrue}
\end{figure}

In deep learning, it is essential to estimate the generalization error of the proposed neural network for understanding its performance. Clearly, the intent is not to compute the generalization error exactly, since it involves an expectation over an unspecified probability distribution. Nevertheless, we can access the test error in the framework of synthetic shot gathers, hence we can use test error minus training error as a good proxy for the generalization error. For the purpose of assessing whether the network can truly generalize ``out of sample" (when the training and testing geophysical models are very different) we train it with the samples collected from the submodels of Marmousi, but test it on the BP 2004 benchmark model (Figure~\ref{fig:bp_vtrue}). With the extrapolated low frequency data predicted by the neural network trained on the submodels of Marmousi, we perform the EFWI-CNN on the BP 2004 benchmark model (Figure~\ref{fig:bp_vtrue}).

\begin{figure}[H]
\centering
\includegraphics[width=0.6\columnwidth]{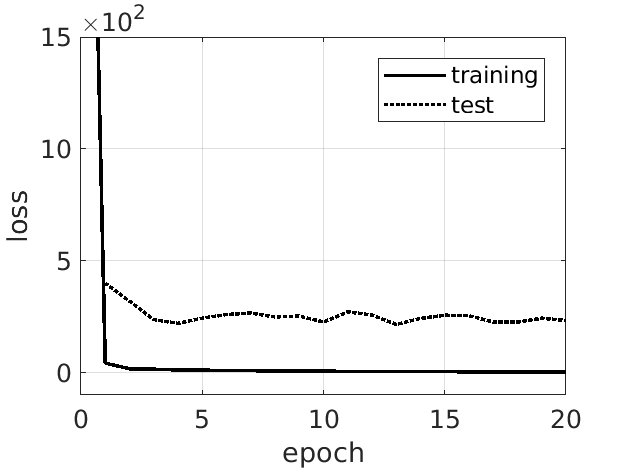}
\caption{The learning curves}
\label{fig:bp_training_loss_band_0_6}
\end{figure}

\begin{figure}[H]
\centering
\includegraphics[width=\columnwidth]{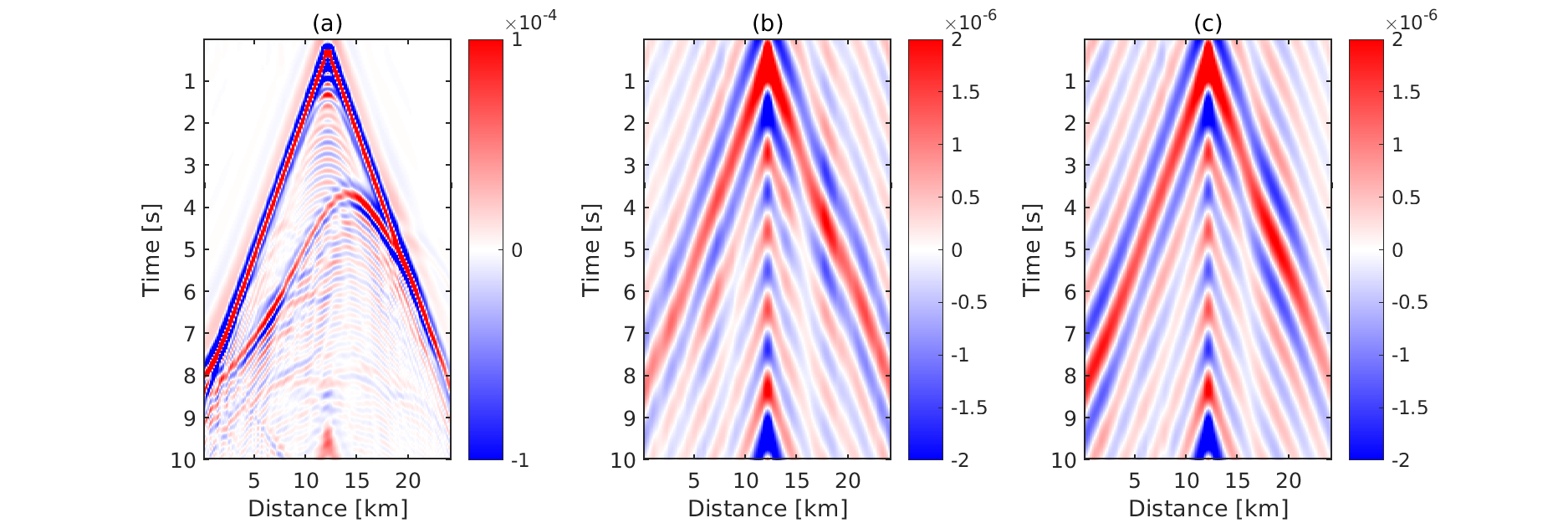}      
\caption{The extrapolation result on the BP model: comparison among the (a) bandlimited recordings ($0.6-20$Hz), (b) predicted and (c) true low frequency recordings ($0.1-0.5$Hz). The neural network trained on the Marmousi2 submodels can recover the low frequencies synthesized from the BP model, which illustrates that the proposed neural network can generalize well.}
\label{fig:bp_records_filter_ishot_15}
\end{figure}

\begin{figure}[H]
\centering
\includegraphics[width=\columnwidth]{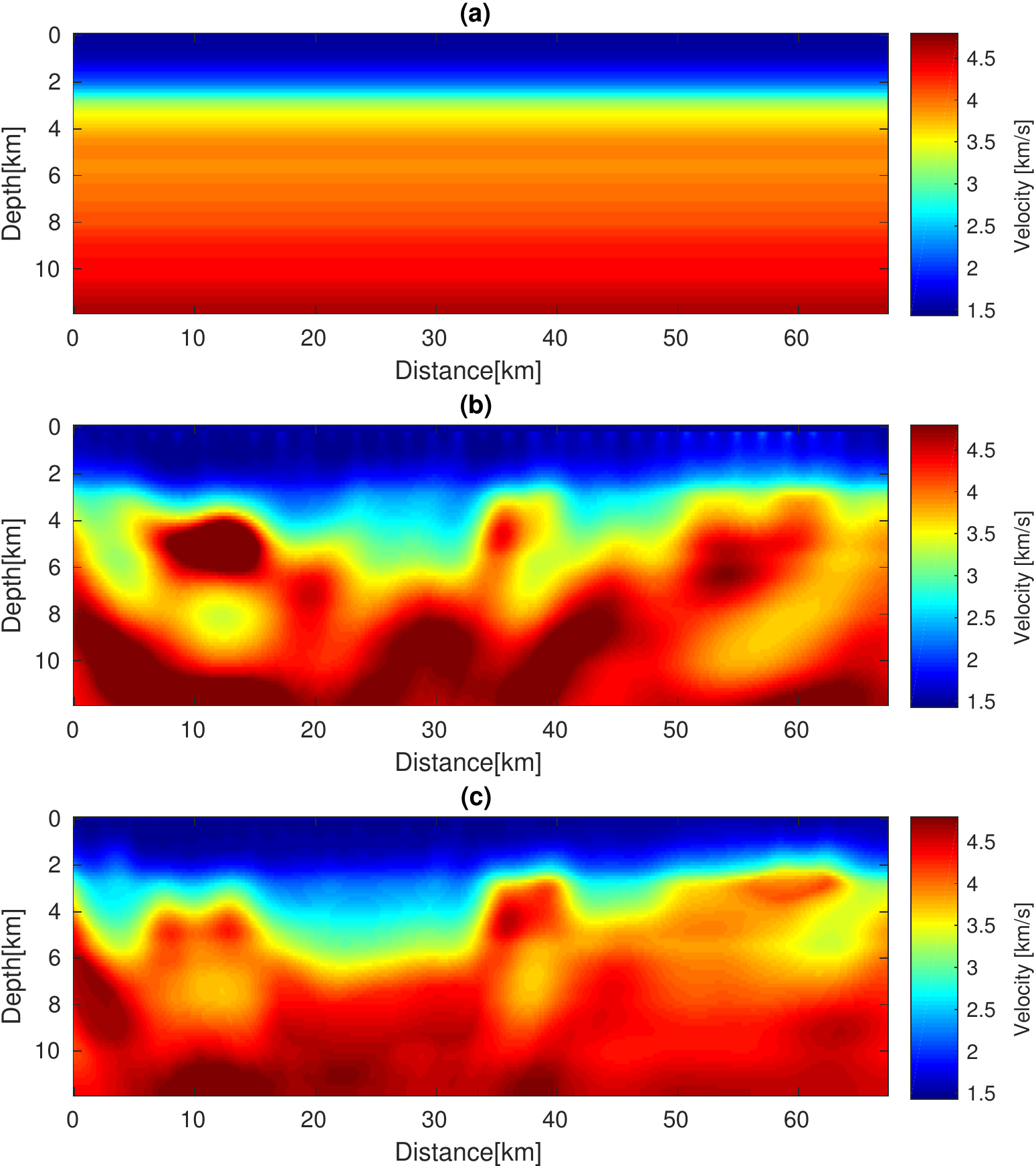}      
\caption{Comparison among (a) the initial model for FWI on the BP model, the inverted low-wavenumber velocity models using (b) 0.3Hz extrapolated data and (c) 0.3Hz true data. The inversion results in (b) and (c) are started from the initial model in (a).}
\label{fig:bp_v0_compare_0_6Hz}
\end{figure}

To reduce the computation burden, we downsample the BP 2004 benchmark model to $80 \times 450$ grid points with a grid interval of 150m. It is challenging for FWI to use only the bandlimited data to invert the shallow salt overhangs and the salt body with steeply dipping flanks in the BP model. Numerical examples show that, starting from the bad initial model (Figure~\ref{fig:bp_v0_compare_0_6Hz}(a)), the highest starting frequency to avoid cycle-skipping on this model is 0.3Hz. Therefore, we should extrapolate the bandlimited data to at least 0.3Hz to invert the BP model successfully.

In this example, we use a 7Hz Ricker wavelet as the source to simulate the fullband seismic records on both the training models (submodels of Marmousi) and the test model (BP model). The sampling interval and the total recording time are 5ms and 10s, respectively. To collect the input of the CNN, a highpass filter where the low frequency band ($0.1-0.5$Hz) is exactly zero is applied to the fullband seismic data. The bandlimited data ($0.6-20$Hz) are fed into the proposed CNN model to extrapolate the low frequency data in $0.1-0.5$Hz trace by trace. 

Figure~\ref{fig:bp_training_loss_band_0_6} shows the learning curves of training process across the 20 epoch. Figure~\ref{fig:bp_records_filter_ishot_15} compares the extrapolation result of one shot gather on the BP model where the shot is located at 31.95km. The neural network can recover the low frequencies of reflections with some degree of accuracy. Even though the information contained in the data collected on Marmousi is physically unlike that of the salt dome model, the pretrained neural network can successfully find an approximation of their low frequencies based on the bandlimited inputs.

The extrapolated low frequency data are used to invert the low-wavenumber velocity model with the conventional FWI method. We observe that the accuracy of extrapolated low frequency decreases as the offset increases, so we limit the maximum offset to 12km. Starting from the initial model (Figure~\ref{fig:bp_v0_compare_0_6Hz}(a)), Figure~\ref{fig:bp_v0_compare_0_6Hz}(b) and Figure~\ref{fig:bp_v0_compare_0_6Hz}(c) show the low-wavenumber inverted models using 0.3Hz extrapolated data and 0.3Hz true data, respectively. Compared to the initial model, the resulting model using the 0.3Hz extrapolated data reveals the positions of the high and low velocity anomalies, which is almost the same as that of true data. The low-wavenumber background velocity models can still initialize the frequency-sweep FWI in the right basin of attraction.

\begin{figure}[H]
\centering
\includegraphics[width=\columnwidth]{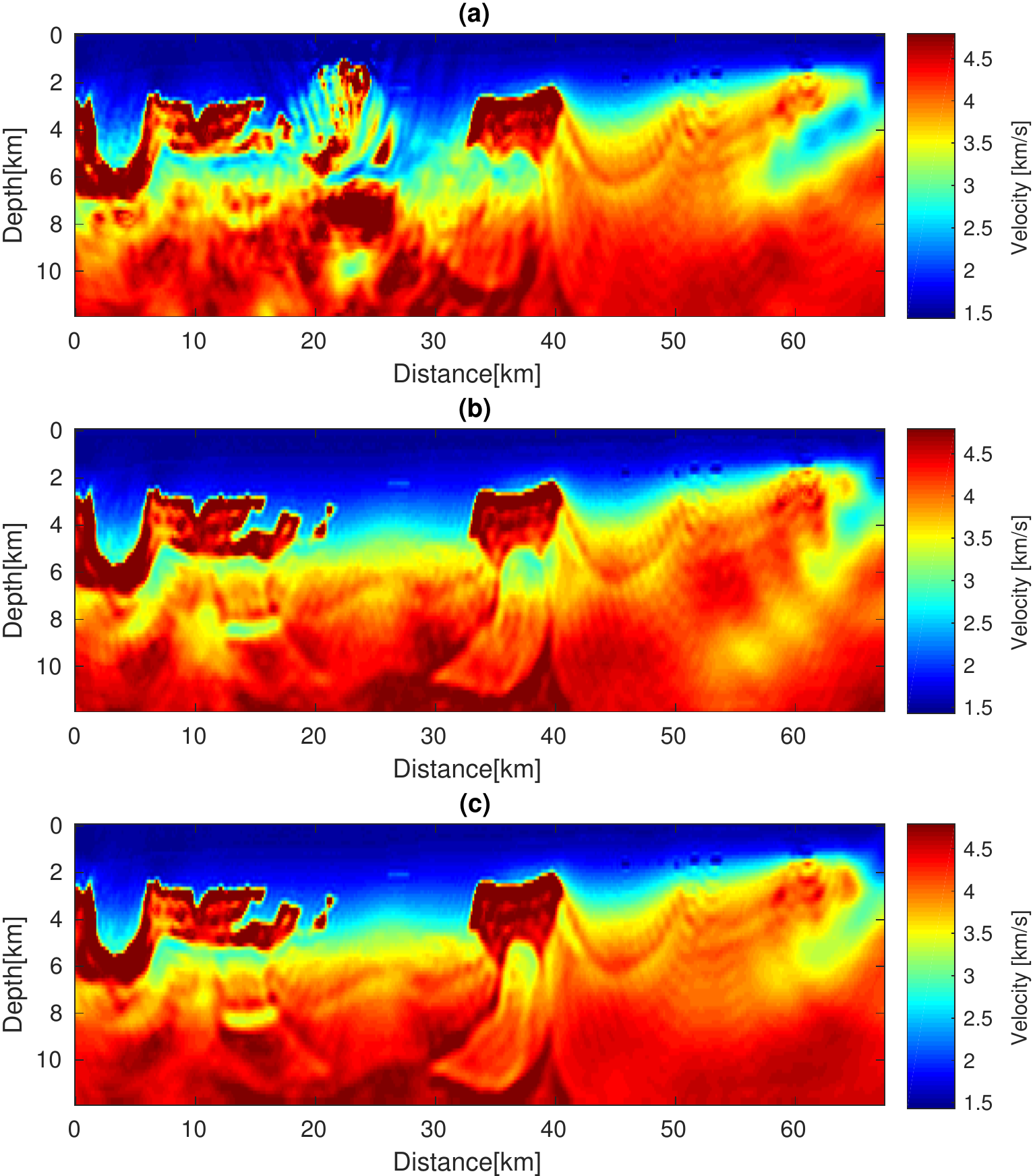}      
\caption{Comparison of the inverted models from FWI using 0.6-2.4Hz bandlimited data. In (a), resulting model starts from the original initial model. In (b), resulting model starts from the inverted low-wavenumber velocity model using 0.3Hz extrapolated data. In (c), resulting model starts from the inverted low-wavenumber velocity model using 0.3Hz true data.}
\label{fig:bp_final_compare_0_6Hz}
\end{figure}

Figure~\ref{fig:bp_final_compare_0_6Hz} compares the inverted models from FWI using 0.6-2.4Hz bandlimited data, starting from the respective low-wavenumber models in the previous figure. In (a), the resulting model starts from the original initial model. In (b), the resulting model starts from the inverted low-wavenumber velocity model using 0.3Hz extrapolated data. In (c), the resulting model starts from the inverted low-wavenumber velocity model using 0.3Hz true data. With the low-wavenumber velocity model, FWI can find the accurate velocity boundaries by exploring higher frequency data. However, the inversion settles in a wrong basin with only the higher frequency components. The low frequencies extrapolated with deep learning are reliable enough to overcome the cycle-skipping problem on the BP model, even though the training data set is ignorant of the particular subsurface structure of BP -- salt bodies. Therefore, the neural network approach has the potential to deal favorably with real field data.

\begin{figure}[H]
\centering
\includegraphics[width=\columnwidth]{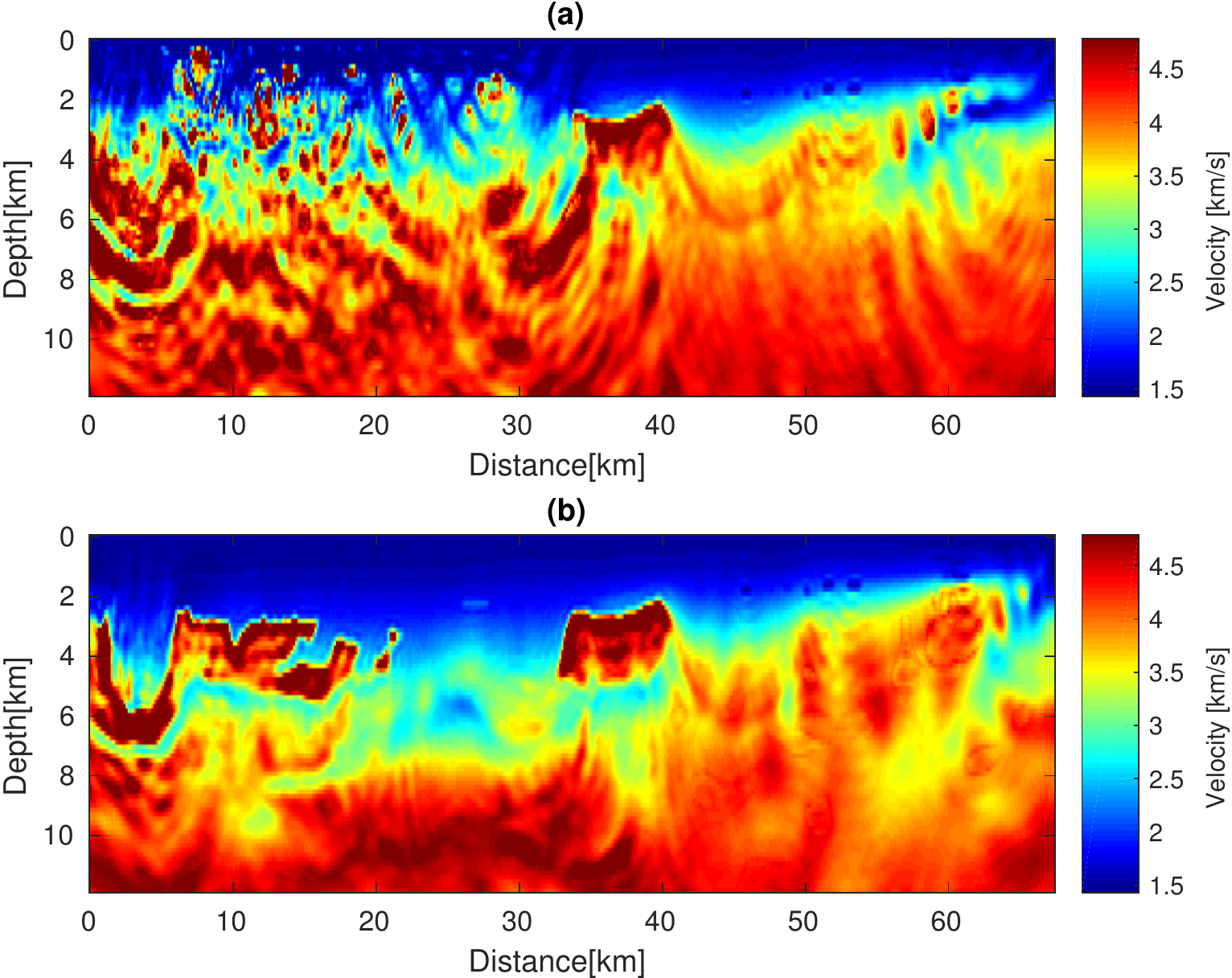}      
\caption{Comparison of the inverted models from FWI using 0.9-2.4Hz bandlimited data. In (a), resulting model starts from the original initial model. In (b), resulting model starts from the inverted low-wavenumber velocity model using 0.3Hz and 0.6Hz extrapolated data. The extrapolated data below 0.9Hz are recovered by 0.9-20Hz bandlimited data.}
\label{fig:bp_final_compare_0_9Hz}
\end{figure}

\begin{figure}[H]
\centering
\includegraphics[width=\columnwidth]{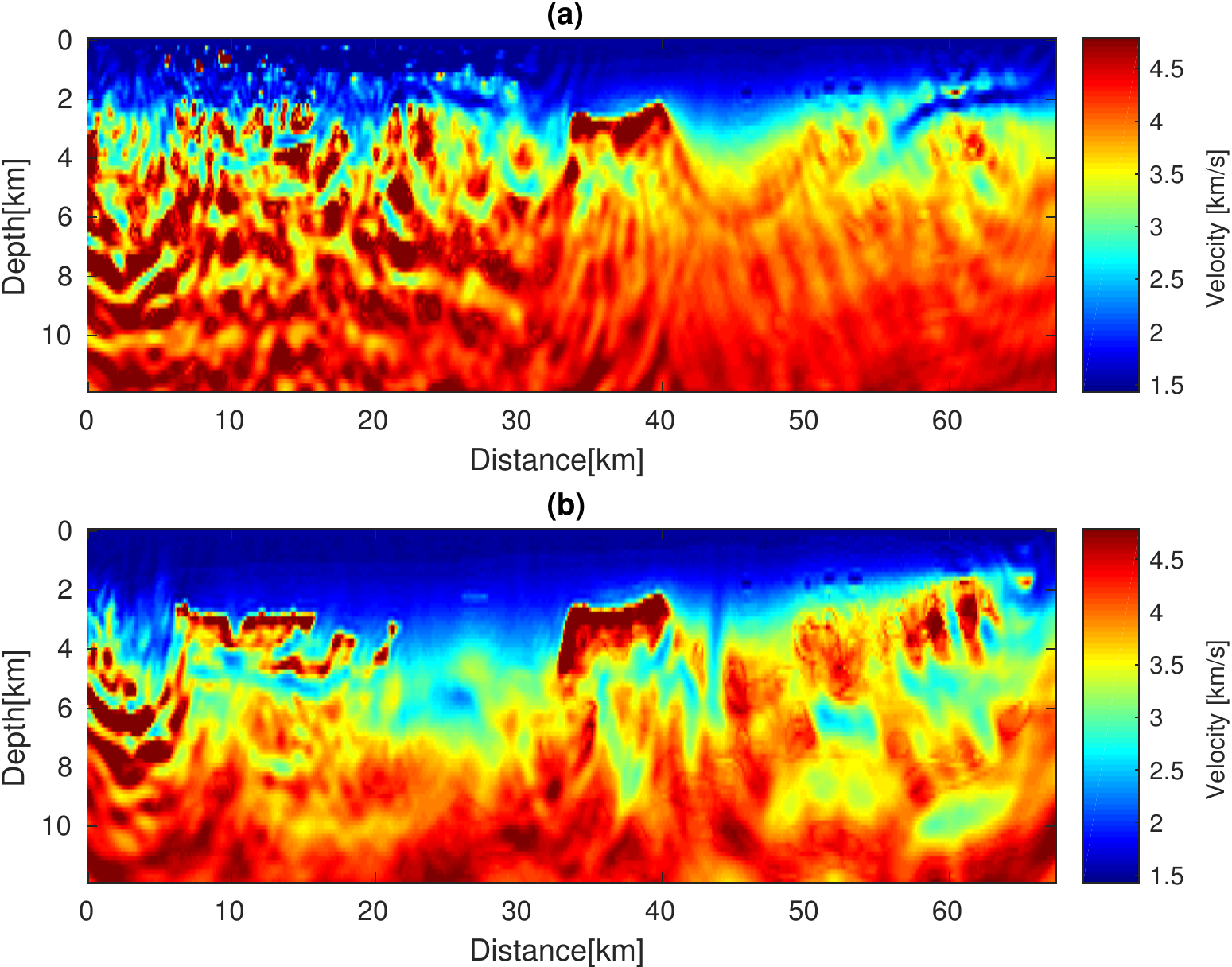}      
\caption{Comparison of the inverted models from FWI using 1.2-2.4Hz bandlimited data. In (a), resulting model starts from the original initial model. In (b), resulting model starts from the inverted low-wavenumber velocity model using 0.3Hz, 0.6Hz and 0.9Hz extrapolated data. The extrapolated data below 1.2Hz are recovered by 1.2-20Hz bandlimited data.}
\label{fig:bp_final_compare_1_2Hz}
\end{figure}

\begin{figure}[H]
\centering
\includegraphics[width=\columnwidth]{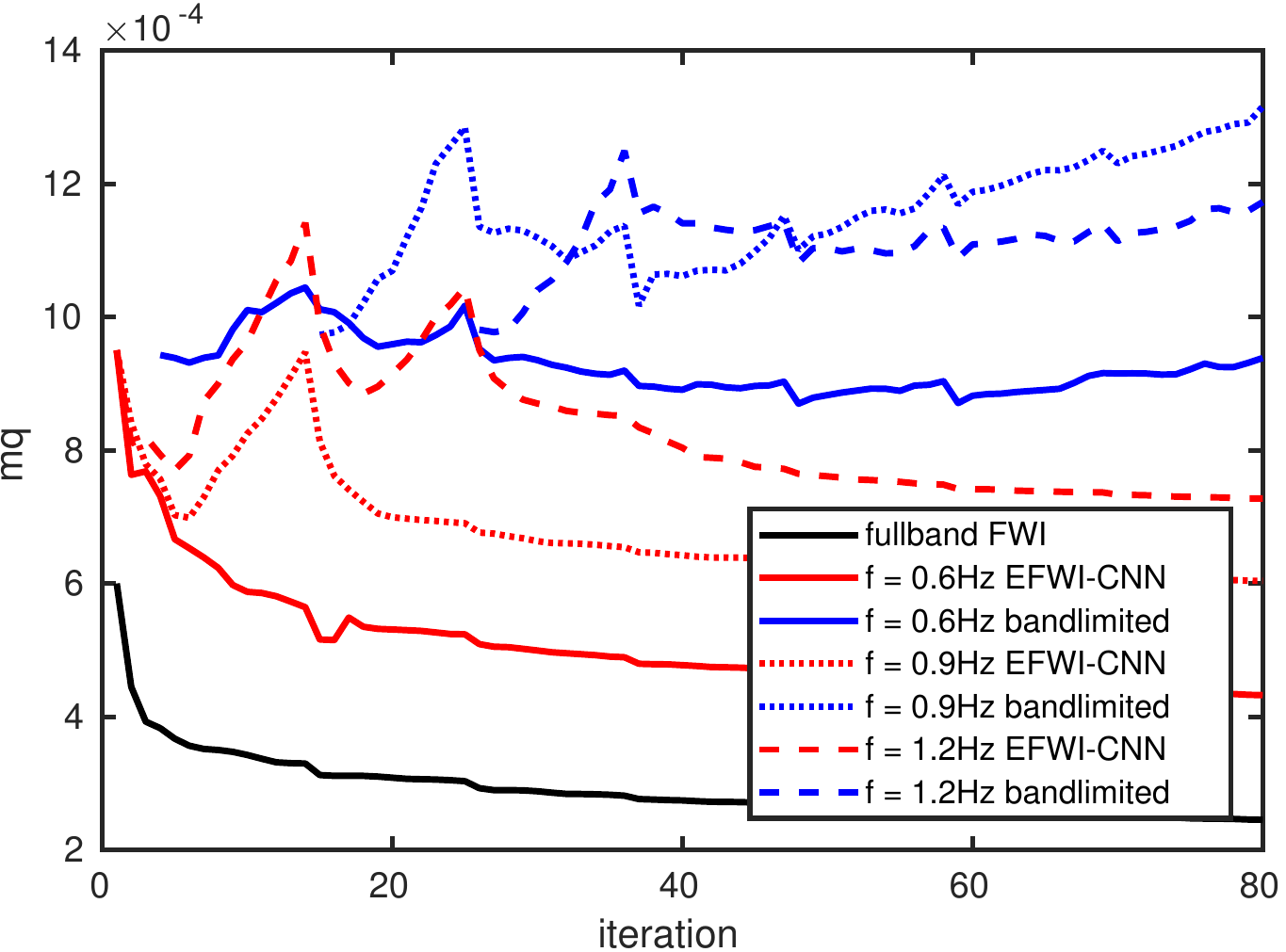}      
\caption{Comparison of the quality of the inverted models at each iteration for FWI using fullband data (black line), EFWI-CNN (red line) and FWI using only bandlimited data (blue line). $f$ donates the lowest frequency of the bandlimited data. The highest frequency of inversion is fixed at 2.4Hz. The performance of EFWI-CNN decreases with the increase of the lowest frequency of the bandlimited data. However, compared with FWI using only bandlimited data, EFWI-CNN improves the quality of the inverted model very well.}
\label{fig:NORM_comparison_iteration}
\end{figure}

So far, the experiments on BP 2004 have assumed that data are available in a band starting at 0.6Hz. We now study the performance of EFWI-CNN when this band starts at a frequency higher than 0.6Hz. We still start the frequency-sweep FWI with 0.3Hz extrapolated data, and the highest frequency is still fixed at 2.4Hz. Figure~\ref{fig:bp_final_compare_0_9Hz} and Figure~\ref{fig:bp_final_compare_1_2Hz} compare the conventional FWI and EFWI-CNN results with data bandlimited above 0.9Hz and 1.2Hz, respectively. With the increase of the lowest frequency of bandlimited data, Figure~\ref{fig:NORM_comparison_iteration} compares the quality of the inverted models at each iteration for FWI using fullband data, EFWI-CNN, and FWI using only the bandlimited data. The norm of the relative model error is used to evaluate the model quality, as \cite[]{brossier2009seismic}
\begin{equation}
mq = \frac{1}{N} \Arrowvert \frac{\textbf{m}_{inv}-\textbf{m}_{true}}{\textbf{m}_{true}} \Arrowvert _2
\end{equation}
where $\textbf{m}_{inv}$ and $\textbf{m}_{true}$ are the inverted model and the true model, respectively. $N$ denotes the number of grid point in the computational domain. The performance of EFWI-CNN of course decreases with the increase of the lowest frequency of the bandlimited data, because this leads to more extrapolated data involved in the frequency-sweep FWI. The more iterations of FWI with the extrapolated data, the more errors the inverted model will have before exploring the true bandlimited data. Overfitting of the unfavorable extrapolated data makes the inversion worse after several iterations with the extrapolated data. However, EFWI-CNN is still superior to using FWI with only bandlimited data. We observe that EFWI-CNN with the current architecture still helps to reduce the inverted model error on the BP model when the lowest available frequency is as high as 1.2Hz.

Finally, we encountered a puzzling numerical phenomenon: the accuracy of the extrapolated data at the single frequency 0.3Hz depends very weakly on the band in which data is available, whether it be $[0.6, 20]$Hz or $[1.2, 20]$Hz for instance. As mentioned earlier, extrapolating data from 0.3Hz to 1.2Hz, so as to be useful for EFWI starting at 1.5Hz, is the much tougher task.

\section{Discussions and Limitations}
The most important limitation of CNN for bandwidth extension is the possibly large generalization error that can result from an incomplete training set, or an architecture unable to predict well out of sample. As a data-driven statistical optimization method, deep learning requires a large number of samples (usually millions) to become an effective predictor. Since the training data set in this example is small but the model capacity (trainable parameters) is large, it is very easy for the neural network to overfit, which seriously deteriorates the extrapolation accuracy. Therefore, in practice, it is standard to use regularization or dropout, with only empirical evidence that this addresses the overfitting problem. 

In addition, the training time for deep learning is highly related to the size of the dataset and the model capacity, and thus is very demanding. To speed up the training by reducing the number of weights of neural networks, we can downsample both the inputs and outputs, and then use bandlimited interpolation method to recover the signal after extrapolation. 

Another limitation of deep learning is due to the unbalanced data. The energy of the direct wave is very strong compared with that of the reflected waves, which biases the neural networks towards fitting the direct wave and contributing less to the reflected waves. Therefore, the extrapolation accuracy of the reflected waves is not as good as that of the primary wave in this example.

\begin{figure}[H]
\centering
\includegraphics[width=\columnwidth]{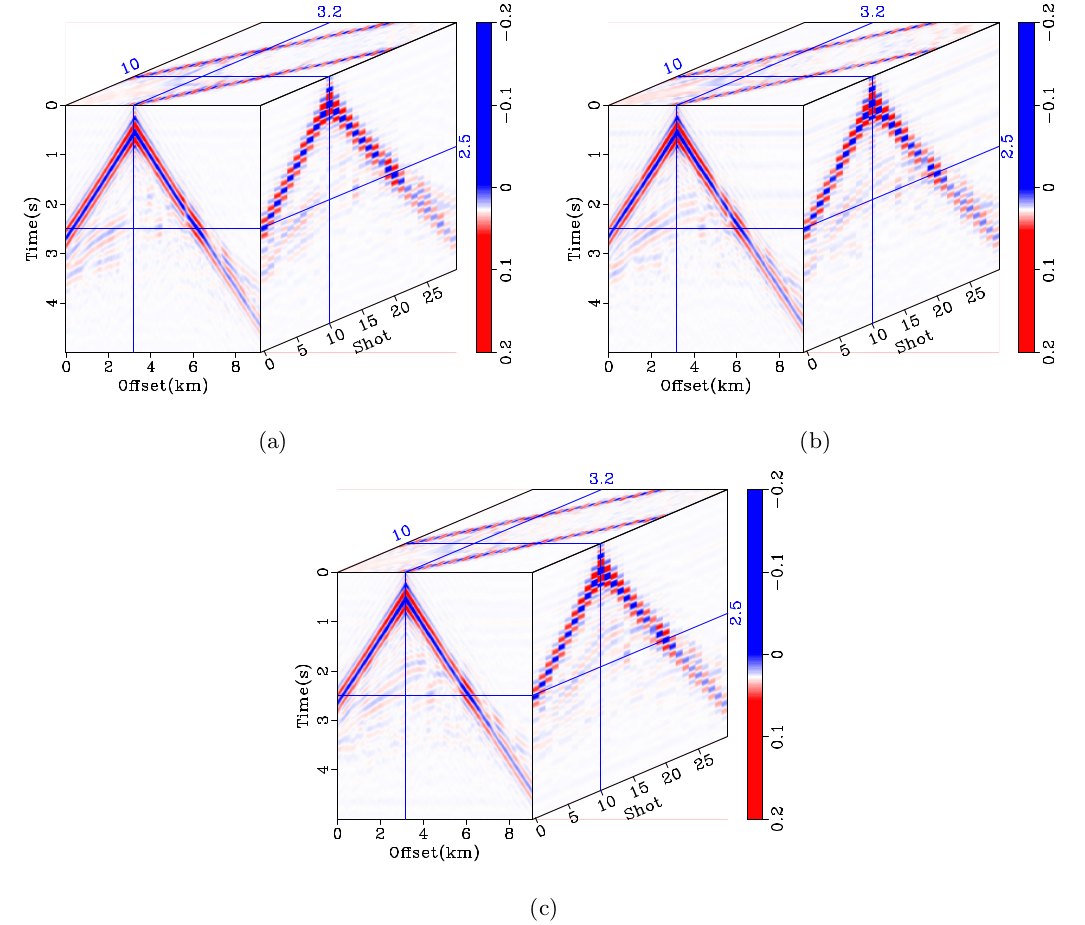}      
\caption{The extrapolation result on the Marmousi model: comparison of extrapolation using (a) one trace, (b) five traces and (c) seven traces as the input of neural network. Multiple-trace extrapolation shows better coherence along traces by leveraging the spatial relationship existing in the input.}
\label{fig:Figure25}
\end{figure}

One limitation of trace-by-trace extrapolation is that the accumulation of the predicted errors reduces the coherence of the event across traces. Hence, multi-trace extrapolation can alleviate this problem to a certain degree by leveraging the spatial relationship existing in the input. The design of the architecture in Figure~\ref{fig:cnn} is flexible to import multiple traces as the input of neural network. To extrapolate the low frequency signal of a single trace, $ntr$ traces in the neighborhood of the single trace have been used as the input of the neural network. Then we only need to change the size of filter on the first convolutional layer from $200\times1$ to $200\times ntr$ and keep the rest the same. Figure~\ref{fig:Figure25} compares the extrapolated low frequency data ($0.1-5Hz$) on the full-size Marmousi model using one trace ($ntr=1$), five traces ($ntr=5$) and seven traces ($ntr=7$) as the input of neural network. The predicted low frequency data using multiple-trace input show better coherence along traces compared to that using trace-by-trace extrapolation. Additionally, more numerical experiments show that multiple-trace extrapolation helps to reduce the random noise but is unhelpful to correlated noise.

Even though we are encouraged by the ability of a CNN to generate $[0.1, 0.5]$Hz data for the BP 2004 model, much work remains to be done to be able to find the right architecture that will generate data in larger frequency bands, for instance in the $[0.1, 1.4]$Hz band. Finding a suitable network architecture, hyperparameters, and training schedule for such cases remains an important open problem. Other community models, and more realistic physics such as elastic waves, are also left to be explored.

Finally, the influence of different physics between the training and test dataset is left to be studied. This point is important for the application to field data. Even though we show the robustness of the proposed method in dealing with uncertainty due to random noise, a different forward modeling operator and a poorly-known source wavelet, more stable neural network and training strategies are yet to be proposed to overcome the challenges of field data, such as both strong correlated and uncorrelated noise, complex and unknown wavelet shot by shot, viscoelasticity and anisotropy.

\section{Conclusions}
In this paper, deep learning is applied to the challenging bandwidth extension problem that is essential for FWI. We formulate bandwidth extension as a regression problem in machine learning and propose an end-to-end trainable model for low frequency extrapolation. Without preprocessing on the input (the bandlimited data) and post-processing on the output (the extrapolated low frequencies), CNN sometimes have the ability to recover the low frequencies of unknown subsurface structure that are completely missing at the training stage. The extrapolated low frequency data can be reliable to invert the low-wavenumber velocity model for initializing FWI on the bandlimited data without cycle-skipping. Even though there is freedom in choosing the architectural parameters of the deep neural network, making the CNN have a large receptive field is necessary for low frequency extrapolation. The extrapolation accuracy can be further modified by adjusting the architecture and hyperparameters of the neural networks depending on the characteristics of the bandlimited data. 

\bibliography{paper}

\end{document}